\documentclass[aps,preprintnumbers,eqsecnum,amsmath,amssymb,showpacs,nofootinbib]{revtex4}
\usepackage{graphicx}% Include figure files
\usepackage{dcolumn}% Align table columns on decimal point
\usepackage{bm}% bold math
\usepackage{epsfig}

\begin{document}
%\preprint{HEPHY-PUB 889/10}
%\preprint{}
\title{Decay constants of heavy pseudoscalar mesons from QCD sum rules}
\author{Wolfgang Lucha$^{a}$, Dmitri Melikhov$^{a,b,c}$, and Silvano Simula$^{d}$}
\affiliation{
$^a$HEPHY, Austrian Academy of Sciences, Nikolsdorfergasse 18, A-1050, Vienna, Austria\\
$^b$SINP, Moscow State University, 119991, Moscow, Russia\\
$^c$Faculty of Physics, University of Vienna, Boltzmanngasse 5, A-1090 Vienna, Austria\\
$^d$INFN, Sezione di Roma III, Via della Vasca Navale 84, I-00146, Roma, Italy}
\date{\today}
\begin{abstract}
We revisit the sum-rule extraction of the decay constants of the
$D$, $D_s$, $B$, and $B_s$ mesons from the two-point correlator of
heavy--light pseudoscalar currents. We use the operator product
expansion of this correlator expressed in terms of the 
$\overline{\rm MS}$ heavy-quark mass, for which the perturbative
expansion exhibits a reasonable convergence. Our main emphasis is laid 
on the control over the uncertainties in the decay constants, 
related both to the input QCD parameters and to the limited
accuracy of the method of sum-rules. This becomes possible due to the 
application of our procedure of extracting hadron observables that involves as novel 
feature dual thresholds depending on the Borel parameter. 
For charmed mesons, we find the decay constants $f_D=(206.2\pm 7.3_{(\rm OPE)}\pm 5.1_{(\rm
syst)})\; {\rm MeV}$ and $f_{D_s}=(245.3\pm 15.7_{(\rm OPE)}\pm
4.5_{(\rm syst)})\; {\rm MeV}$. For beauty mesons, the decay
constants turn out to~be extremely sensitive to the precise value
of $\overline{m}_b(\overline{m}_b)$. By requiring our sum-rule
estimate to match~the average of the lattice results for $f_B$, a very accurate value 
$\overline{m}_b(\overline{m}_b)=4.245\pm
0.025\; {\rm GeV}$ is extracted, leading to $f_{B} = (193.4
\pm 12.3_{\rm (OPE)} \pm 4.3_{\rm (syst)})\; {\rm MeV}$ and
$f_{B_s} = (232.5 \pm 18.6_{\rm (OPE)} \pm 2.4_{\rm (syst)})\;
{\rm MeV}$.
\end{abstract}
\maketitle

\section{Introduction}
The extraction of the ground-state decay constant within the
method of QCD sum rules \cite{svz} is a complicated problem:

First, one should derive a reliable operator product expansion
(OPE) for the Borel-transformed (vulgo ``Borelized'') correlator
$\Pi(\tau)$. The OPE in QCD is a double expansion: a perturbative
expansion in powers of the strong coupling~$\alpha_s$ and an
expansion~in powers of the Borel parameter $\tau$ in terms of
condensates of increasing dimension. In practice,~one has a
truncated double series for which at most a few lowest-order terms
can be calculated. This truncated series~for~the correlator
depends explicitly on the renormalization scheme and scale (even
in those cases where the full correlator~does not), and the
magnitude of the unknown higher-order terms crucially depends on
the relevant choice of this scheme~and scale. Therefore,
controlling higher-order perturbative corrections poses a serious
problem.

Since the pioneering work \cite{aliev}, the correlator expressed
in terms of the on-shell (or pole) heavy-quark mass has~been
employed for the extraction of the decay constant. However, after
the three-loop result for the correlator has appeared
\cite{chetyrkin} it became evident that the perturbative expansion
in terms of the on-shell mass shows no signal of convergence:~the
LO, NLO, and NNLO terms all give comparable contributions to the
decay constant. In contrast to this, a reshuffling~of the
perturbative expansion making use of the $\overline{\rm MS}$ mass
of the heavy quark leads to a clear hierarchy of the perturbative
contributions to the decay constant \cite{jamin}.\footnote{As
rather unpleasant consequence, the decay constant extracted from
the three-loop correlator in terms of the on-shell mass
turns~out~to be {\em considerably\/} smaller than the estimate
obtained from the three-loop correlator in terms of the running
$\overline{\rm MS}$ mass \cite{lms}. For the translation from one
scheme to the other one employs the three-loop relation between
the on-shell and the running quark masses.} Following
\cite{jamin}, we adopt the OPE of the relevant correlator in terms
of the~$\overline{\rm MS}$ heavy-quark mass, denoted hereafter by
$m_Q\equiv \overline{m}_Q(\overline{m}_Q)$. Therefore, the
on-shell mass does not appear in our analysis.

Second, the knowledge of the correlator for only moderate values
of $\tau$ allows one to extract the characteristics of the bound
state with some error which reflects the intrinsic uncertainty of
the method of QCD sum-rules. Gaining~control over this systematic
uncertainty is a rather subtle problem, as it has been shown in
\cite{lms_2ptsr}. Moreover, since higher~multiloop perturbative
calculations are becoming available and the knowledge of the
fundamental QCD parameters is improving, the accuracy of the OPE
of the relevant correlators is increasing. Therefore, the
intrinsic systematic uncertainty of~the QCD sum-rule method may
become competitive with the decreased OPE uncertainties, as we
shall show to be the~case for the $D$-meson decay constant $f_D$.

This work presents a detailed analysis of the decay constants of
the $D_{(s)}$ and $B_{(s)}$ mesons, with emphasis on~acquiring
control over all the uncertainties --- of both OPE and intrinsic
(i.e., systematic) origin --- in these quantities. Recently, we
formulated a novel algorithm for extracting bound-state parameters
from OPEs for the correlators which opens the possibility to
arrive at realistic error estimates for the extracted hadron
parameters \cite{lms_new}. The efficiency of our~algorithm has
been established in potential models: there the exact ground-state
decay constants may be computed by solving~the Schr\"odinger
equation. Moreover, it has been explicitly demonstrated that the
{\it extraction procedures\/} of the ground-state parameters in
QCD and in potential models are very close to each other
quantitatively as soon as the (quark--hadron) duality is
implemented in both theories in the same way \cite{lms_qcdvsqm}.

This paper is organized as follows: Section \ref{Sec:CSR}
summarizes existing results on the OPEs for pseudoscalar
heavy-light two-point functions and presents details of our
algorithm for extracting the ground-state contribution to
this~correlator. Section \ref{Sec:DDC} sketches our analysis of
$f_{D}$ and $f_{D_s}$, testing and proving the efficiency of our
formalism. Section \ref{Sec:BDC} provides the corresponding
analysis of $f_{B}$ and $f_{B_s}$. Section \ref{Sec:S&C}
summarizes our conclusions.

\section{Correlator and sum rule}\label{Sec:CSR}
We consider the correlator
\begin{eqnarray}
\label{Pi_QCD} \Pi(p^2)=i \int d^4x\, e^{ipx}\langle
0|T\left(j_5(x)j^\dagger_5(0)\right)| 0\rangle
\end{eqnarray}
of two pseudoscalar heavy--light currents
\begin{eqnarray}
j_5(x)=(m_Q+m)\bar q(x) i\gamma_5 Q(x).
\end{eqnarray}
The Borel-transformed OPE series for this correlator is of the
form
\begin{eqnarray}
\label{OPE_QCD}
\Pi(\tau)=\int\limits^\infty_{(m_Q+m)^2}ds\,e^{-s\tau}\rho_{\rm
pert}(s,\mu) + \Pi_{\rm power}(\tau,\mu),
\end{eqnarray}
where the perturbative spectral density may be found as a power
series in the strong coupling $\alpha_s$:
\begin{eqnarray}
\label{rhopert} \rho_{\rm
pert}(s,\mu)=\rho^{(0)}(s)+\frac{\alpha_s(\mu)}{\pi}\rho^{(1)}(s)+
\left(\frac{\alpha_s(\mu)}{\pi}\right)^2\rho^{(2)}(s)+\cdots.
\end{eqnarray}
The Borel-transformed correlator (\ref{OPE_QCD}) does not depend
on the renormalization scale $\mu$; however, both the perturbative
expansion truncated to some fixed order in $\alpha_s$ and the
truncated power corrections depend on $\mu$. Moreover, the
relative magnitudes of the lowest-order contributions strongly
depend on the choice of the renormalization scheme and/or scale.

A crucial issue for the reliability of a {\em truncated\/}
perturbative expansion is the magnitude of the unknown
higher-order corrections. One can play around with the choice of
the renormalization scale in order to obtain some
properties~of~the known terms of the perturbative expansion. For
example, one may choose the scale $\mu$ by minimizing the
highest-order known correction or by requiring some hierarchy of
the known perturbative contributions. Unfortunately, even~a~clear
hierarchy of several lowest-order perturbative corrections does
not mean that the subsequent corrections are also small. Very
often, the variation of the renormalization scale $\mu$ in some
range is used as an attempt to probe the~magnitude~of the unknown
higher-order corrections. We shall pursue this strategy too,
although there seems to be no rigorous~way~to estimate the size of
these corrections without explicitly calculating them.

In all expressions below, the quark masses $m_Q$ and $m$ and the
strong coupling $\alpha_s$ denote the respective $\overline{\rm
MS}$ running quantities at the scale $\mu$. Note that
$\rho^{(0)}(s)$ depends on $\mu$ implicitly through the quark
masses, whereas all higher-order spectral densities
$\rho^{(n)}(s)$, $n\ge1$, depend on $\mu$ implicitly through the
quark masses and contain, in addition, explicitly $\mu$-dependent
logarithmic terms. Both perturbative spectral density and power
corrections are given below for $\mu=m_Q$.

\subsection{Perturbative spectral density}
The leading-order spectral density is well-known:
\begin{eqnarray}
\label{rho0}
\rho^{(0)}(s)=\frac{N_c}{8\pi^2}(m_Q+m)^2\frac{s-(m_Q-m)^2}{s}
\sqrt{\left[s-(m_Q-m)^2\right]\left[s-(m_Q+m)^2\right]}.
\end{eqnarray}
For the spectral density of order $\alpha_s$ we make use of the
first two terms of its expansion in small mass $m$ derived
in~\cite{jamin}:\footnote{Note that the $O(\alpha_s m^2)$
corrections to $\Pi(\tau)$ cannot be obtained by expanding the
spectral density $\rho^{(1)}$ in powers of $m$: Starting from~the
order $m^3$, the functions $\rho^{(1)}_{m^3}$ etc.\ contain poles
of increasing orders at $s=m_Q^2$ (see Eqs.~(C3) and (C4) of
Ref.~\cite{jamin}). Therefore, after the $s$-integration all these
terms yield contributions of the same order $O(\alpha_s m^2)$ to
the spectral function. For the same reason, the expansion of
$\rho^{(0)}(s)$ in powers of $m$ does not allow one to get the
terms of order $m^3$ in $\Pi(\tau)$. We therefore use the exact
expression for the spectral density $\rho^{(0)}$ instead of
expanding it in powers of $m$.}
\begin{eqnarray}
\rho^{(1)}(s)&=&\rho^{(1)}_{m^0}(s)+\rho^{(1)}_{m}(s)+O(m^2),
\nonumber\\ \rho^{(1)}_{m^0}(s)&=&\frac{N_c}{16\pi^2}C_F
(m_Q+m)^2s(1-x) \left\{(1-x) \left[\,4L_2(x)+2\ln x\ln(1-x)
-(5-2x)\ln(1-x)\right]\right.\nonumber\\ &&\hspace{4.27cm}\left.+
(1-2x)(3-x)\ln x + (17-33x)/2\right\}, \nonumber\\
\rho^{(1)}_{m}(s)&=& \frac{N_c}{8\pi^2}C_F(m_Q+m)^2 m_Q m
\left\{(1-x)\left[4L_2(x)+2\ln x\ln(1-x) -
2(4-x)\ln(1-x)\right]\right.\nonumber\\ &&\hspace{3.72cm}\left. +
2(3-5x+x^2)\ln x + 2(7-9x)\right\},
\end{eqnarray}
where $x\equiv m_Q^2/s$, $N_c=3$, and $C_F=(N_c^2-1)/2N_c$. The
order-$\alpha_s^2$ spectral density reads
\begin{eqnarray}
\rho^{(2)}(s)&=&
R^{(2),s}+\Delta_1\rho^{(2)}+\Delta_2\rho^{(2)}+O(m).
\end{eqnarray}
Here, $R^{(2),s}$ is the spectral function defined by Eq.~(8) of
Ref.~\cite{chetyrkin}, which is provided by the authors through
the~publicly available program $rvs.m$. The authors of
\cite{chetyrkin} have calculated the spectral function $\rho_{\rm
pert}(s)$ for the case $m=0$ in terms of the heavy-quark on-shell
mass. Rewriting the $O(1)$ and $O(\alpha_s)$ spectral densities
$R^{(0),s}$ and $R^{(1),s}$ of \cite{chetyrkin} in terms~of the
running mass generates the corrections $\Delta_1\rho^{(2)}$ and
$\Delta_2\rho^{(2)}$ to the spectral density $\rho^{(2)}(s)$. The
explicit expressions for these corrections are given by Eqs.~(14)
and (15) of Ref.~\cite{jamin} and will not be reproduced here.

\subsection{Power corrections}
For the power corrections we also make use of the expression from
\cite{jamin}:
\begin{eqnarray}
\label{power_QCD} &&\Pi_{\rm power}(\tau,\mu=m_Q)=
(m_Q+m)^2e^{-m_Q^2\tau} \\ \nonumber
&&\hspace{1em}\times\left\{-m_Q\langle \bar qq\rangle \left[
1+\frac{2C_F\alpha_s}{\pi}\left(1-\frac{m_Q^2\tau}2\right)
-\frac{m}{2m_Q}(1+m_Q^2\tau)+\frac{m^2}{2} m_Q^2\tau^2
+\frac{m_0^2\tau}{2}\left(1-\frac{m_Q^2\tau}{2}\right) \right]+
\frac{1}{12}\left\langle{\frac{\alpha_s}{\pi} GG}\right\rangle
\right\}.
\end{eqnarray}
The parameter $m_0^2$ describes the dimension-5 mixed quark--gluon
condensate \cite{jamin}. It is worth noticing that the radiative
corrections to the condensates increase rather fast with the Borel
parameter $\tau$.

In summary, we make use of the expressions for $\Pi(\tau)$ from
\cite{jamin} with minor modifications:

(i) We adopt the ``natural'' threshold $(m_Q+m)^2$ in the spectral
representation for $\Pi(\tau)$ and therefore encounter~only
running $\overline{\rm MS}$ masses in our formulas.

(ii) For $\rho^{(0)}(s)$ we use the exact expression without
performing an expansion in powers of $m$, and for $\rho^{(1)}(s)$
we do not include terms of order $m^2$ and higher.

The OPE parameters required for our analysis are
\begin{eqnarray}
&&m(2\;{\rm GeV})= (3.5 \pm 0.5)\; {\rm MeV},\quad m_s(2\;{\rm
GeV})= (100 \pm 10)\; {\rm MeV},\nonumber\\ &&\langle \bar
qq\rangle(2\; {\rm GeV})=-((267\pm 17) \;{\rm
MeV})^3,\quad\frac{\langle \bar ss\rangle(2\; {\rm GeV})}{\langle
\bar qq\rangle(2\; {\rm GeV})}=0.8\pm 0.3,\quad
\left\langle\frac{\alpha_s}{\pi}GG\right\rangle=(0.024\pm
0.012)\;{\rm GeV}^4,\nonumber\\ && m_0^2=(0.8\pm 0.2)\;{\rm
GeV}^2,\quad\alpha_S(M_Z)=0.1176\pm 0.0020.\label{Eq:OPEP}
\end{eqnarray}
We perform the calculations for two sets of $c$- and $b$-quark
masses $m_Q\equiv \overline{m}_Q(\overline{m}_Q)$: the values from
PDG \cite{pdg}
\begin{equation}m_c=\left(1.27^{+0.07}_{-0.11}\right){\rm GeV},\quad
m_b=\left(4.2^{+0.17}_{-0.07}\right){\rm
GeV},\label{Eq:mc1}\end{equation}and the very accurate values
reported recently in \cite{mb}
\begin{equation}m_c=(1.279\pm 0.013)\;{\rm GeV},\quad m_b=(4.163\pm
0.016)\;{\rm GeV}.\label{Eq:mc2}\end{equation}

\subsection{Sum rule}
The correlator (\ref{Pi_QCD}) may be evaluated by inserting a
complete set of hadronic intermediate states:
\begin{eqnarray}
\label{hadron} \Pi(\tau)={f_Q^2
M_Q^4}e^{-M_Q^2\tau}+\mbox{contributions of higher states},
\end{eqnarray}
where $f_Q$ is the decay constant of the $P_Q$ meson, defined
according to
\begin{eqnarray}
\label{decay_constant} (m_Q+m) \langle 0 |\bar u i\gamma_5 Q| P_Q
\rangle = f_Q M_Q^2.
\end{eqnarray}
For large values of $\tau$, the contributions of the excited
states decrease faster than the ground-state contribution and~the
correlator $\Pi(\tau)$ is dominated by the ground state.
Unfortunately, the truncated OPE does not allow us to
calculate~the correlator at sufficiently large $\tau$, such that
the excited states give a sizable contribution to $\Pi(\tau)$.

According to the quark--hadron duality assumption, the
contributions of excited states and continuum are described by the
QCD perturbative contribution above an effective continuum
threshold $s_{\rm eff}$. This leads to the following relation:
\begin{eqnarray}
\label{SR_QCD} f_Q^2 M_Q^4 e^{-M_Q^2\tau}=\Pi_{\rm dual}(\tau,
s_{\rm eff}(\tau)) \equiv \int\limits^{s_{\rm
eff}(\tau)}_{(m_Q+m)^2}ds\, e^{-s\tau}\rho_{\rm pert}(s,\mu) +
\Pi_{\rm power}(\tau,\mu).
\end{eqnarray}
In the region near the physical continuum threshold at
$s=(M_{V_Q}+m_\pi)^2$, $V_Q$ being the lightest vector meson
containing a quark $Q$, the QCD perturbative spectral density and
the hadron spectral density are rather different. Consequently,
the effective continuum threshold as defined by (\ref{SR_QCD})
turns out to be necessarily a function of the Borel parameter
$\tau$.

We introduce the dual invariant mass $M_{\rm dual}$ and the dual
decay constant $f_{\rm dual}$ by the definitions
\begin{eqnarray}
\label{mdual} M_{\rm dual}^2(\tau)&\equiv&-\frac{d}{d\tau}\log
\Pi_{\rm dual}(\tau, s_{\rm eff}(\tau)),
%\nonumber
\\
\label{fdual} f_{\rm dual}^2(\tau)&\equiv&M_Q^{-4}
e^{M_Q^2\tau}\Pi_{\rm dual}(\tau, s_{\rm eff}(\tau)).
\end{eqnarray}
Notice that the deviation of the dual mass from the actual
ground-state mass provides an indication of the excited-state
contributions picked up by the dual correlator.

In order to determine the decay constant $f_Q$ of the $P_Q$ meson
from the OPE we must execute the following two~steps.

\subsubsection{Borel window}
First, we have to fix the working $\tau$-window where, on the one
hand, the OPE yields a sufficiently accurate description of the
exact correlator (that is, the higher-order radiative and power
corrections are small) and, on the other hand,~the ground state
gives a ``sizable'' contribution to the correlator. Since the
radiative corrections to the condensates~increase rather fast with
$\tau$, it is preferable to stay at the lowest possible values of
$\tau$. We shall therefore fix the Borel window~by the following
criteria: (a) In the window, the power corrections $\Pi_{\rm
power}(\tau)$ should not exceed 30\% of the cut perturbative
correlator $\Pi_{\rm pert}(\tau,s_0)$; this gives the upper
boundary of the $\tau$-window. The ground-state contribution to
the correlator at such $\tau$ values comprises about 50\% of the
correlator. (b) The lower boundary of the $\tau$-window is defined
by requiring the ground-state contribution not to drop below 10\%.
In quantum physics, our algorithm was shown to provide a~good,
even excellent extraction of the ground-state decay constant for
Borel windows determined by these requirements~\cite{lms_2ptsr,
lms_new}.

\subsubsection{Effective continuum threshold}
Second, we must formulate our criterion for the determination of
$s_{\rm eff}(\tau)$. We consider an algorithm for the extraction
of $f_Q$ which takes advantage of the knowledge of the $P_Q$-meson
mass $M_Q$. Our algorithm, constructed in previous~works and, in
quantum-theoretical potential models, proven to work well for
various correlators, is rather simple: We consider a set of
$\tau$-dependent Ans\"atze for the effective continuum threshold,
for simplicity assumed to be all of polynomial~form:
\begin{eqnarray}
\label{zeff} s^{(n)}_{\rm eff}(\tau)=
\sum\limits_{j=0}^{n}s_j^{(n)}\tau^{j}.
\end{eqnarray}
We determine the parameters on the r.h.s.\ of (\ref{zeff}) as
follows: We calculate the dual mass squared according to
(\ref{mdual}) for the $\tau$-dependent $s_{\rm eff}$ of
Eq.~(\ref{zeff}). We then compute $M^2_{\rm dual}(\tau)$ at
several values of $\tau = \tau_i$ ($i = 1,\dots, N$,~where~$N$~can
be taken arbitrary large) chosen uniformly over the Borel window.
Finally, we minimize the squared difference between $M^2_{\rm
dual}$ and the known value $M^2_Q$:
\begin{eqnarray}
\label{chisq}
\chi^2 \equiv \frac{1}{N} \sum_{i = 1}^{N} \left[ M^2_{\rm dual}(\tau_i) - M_Q^2 \right]^2.
\end{eqnarray}
This pins down the parameters of the effective continuum
threshold. As soon as the latter is fixed, it is straightforward
to calculate the decay constant $f_Q$.

According to our recent findings, allowing for a $\tau$-dependence
of the effective threshold leads to visible improvements compared
with the traditional assumption of a $\tau$-independent quantity:
The former yields a much better stability~of~the dual mass
calculated from the dual correlator and allows one to work at
smaller values of $\tau$, where the impact of~power corrections is
reduced.

\subsubsection{Uncertainties of the extracted decay constant}
The above discussion implies that the extracted result for the
decay constant is sensitive both to the precise values~of the OPE
parameters and to the particular prescription for fixing the
effective continuum threshold. The corresponding uncertainties of
the predicted decay constant are labeled as its {\it OPE-related
error\/} and its {\it systematic~error}, respectively:
\begin{enumerate}\item{\it OPE-related error}. The OPE-related
uncertainty is estimated as follows: We perform a bootstrap
analysis~\cite{bootstrap}~by allowing the OPE parameters to vary
over the ranges quoted in Eqs.~(\ref{Eq:OPEP}), using 1000
bootstrap events. Gaussian distributions for all parameters but
$\mu$ are employed. For $\mu$ we assume a uniform distribution in
the corresponding range, which we choose to be $1 \leq
\mu\;(\mathrm{GeV}) \leq 3$ for charmed mesons and $2 \leq
\mu\;(\mathrm{GeV}) \leq 8$ for beauty mesons.~The resulting
distribution of the decay constant turns out to be close to a
Gaussian shape. The quoted OPE-related error is therefore the
Gaussian error.\item{\it Systematic error}. The systematic
uncertainty of some hadron parameter obtained by the sum-rule
method (i.e., the error related to the intrinsically limited
accuracy of this approach) represents the perhaps most subtle
point in all applications of this method. So far no way to provide
a {\it rigorous\/} --- in the mathematical sense --- systematic
error has been devised. However, a {\it realistic} estimate of the
corresponding error may be found:~As~prompted by detailed
comparisons of the extraction of the decay constant in QCD and in
potential models \cite{lms_qcdvsqm}, the band~of~$f_P$ values
spanned by the linear, quadratic, and cubic Ans\"atze for the
effective continuum threshold contains~the~true value of the decay
constant. Trusting in these findings, the half-width of this band
is interpreted as the systematic error of the decay constant.
Presently, we do not envisage any other possibility to provide
more reliable estimates for the systematic error.\end{enumerate}

\section{\boldmath Decay constants of the $D$ and $D_s$ mesons}
\label{Sec:DDC}
\subsection{Decay constant of the $D$ meson}
\begin{figure}[!ht]
\begin{tabular}{ccc}
\includegraphics[width=5.75cm]{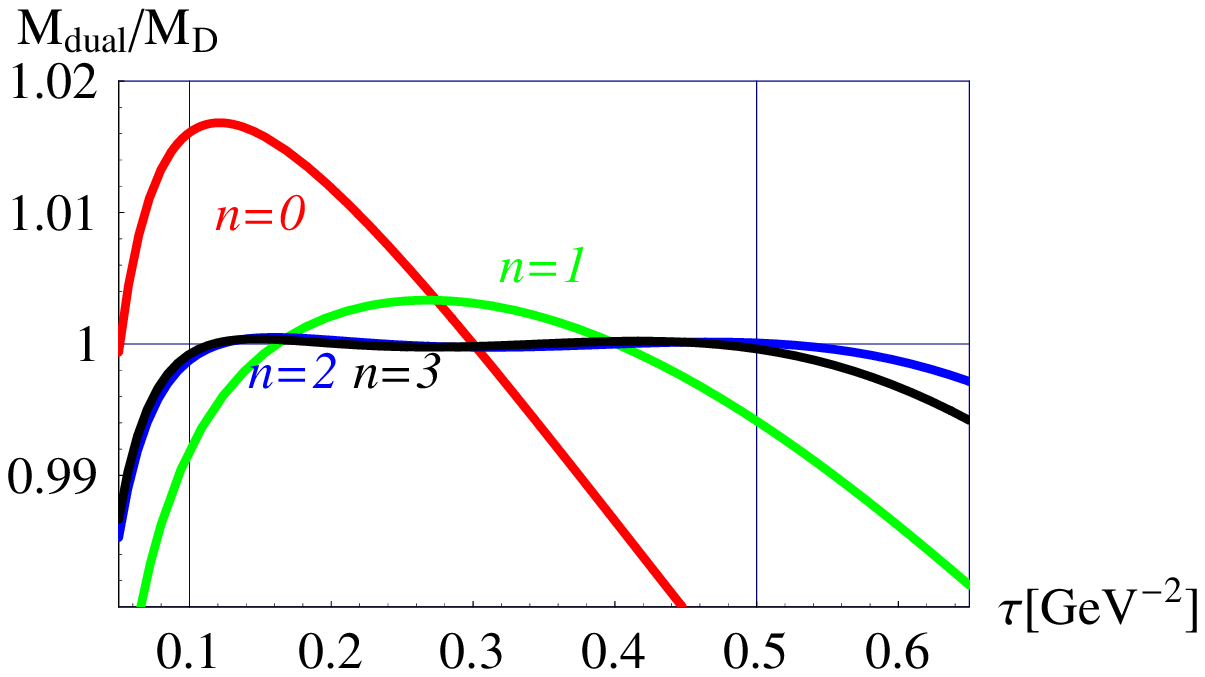}&
\includegraphics[width=5.75cm]{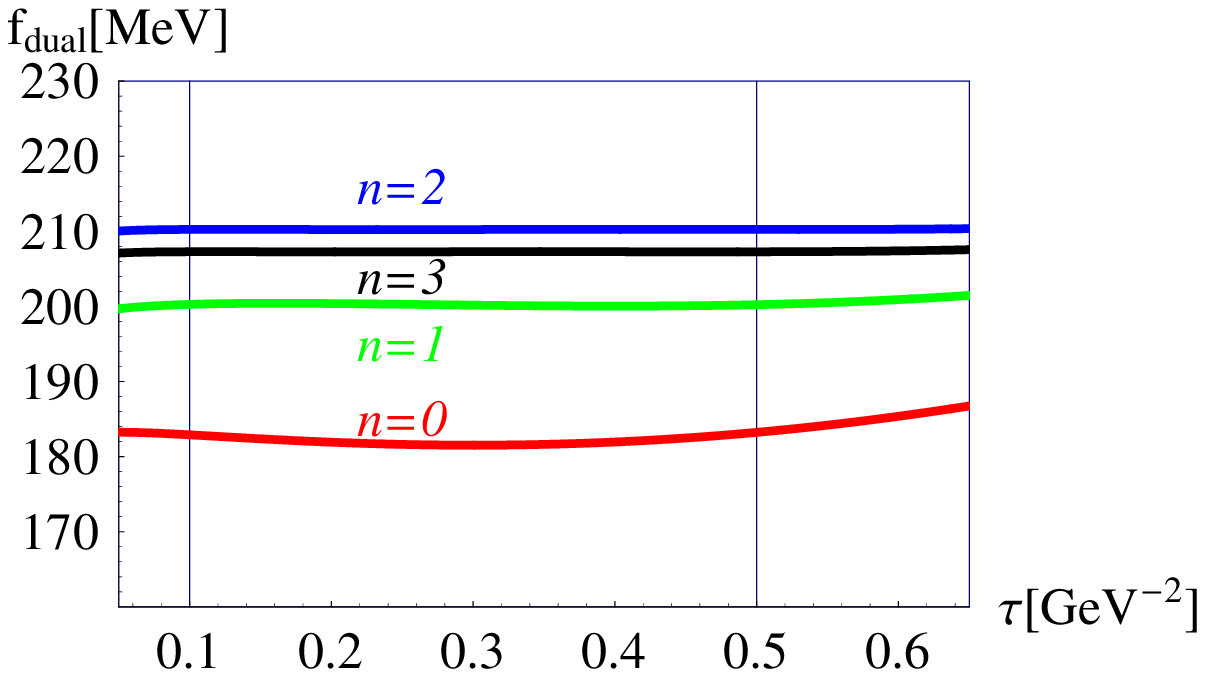}&
\includegraphics[width=5.75cm]{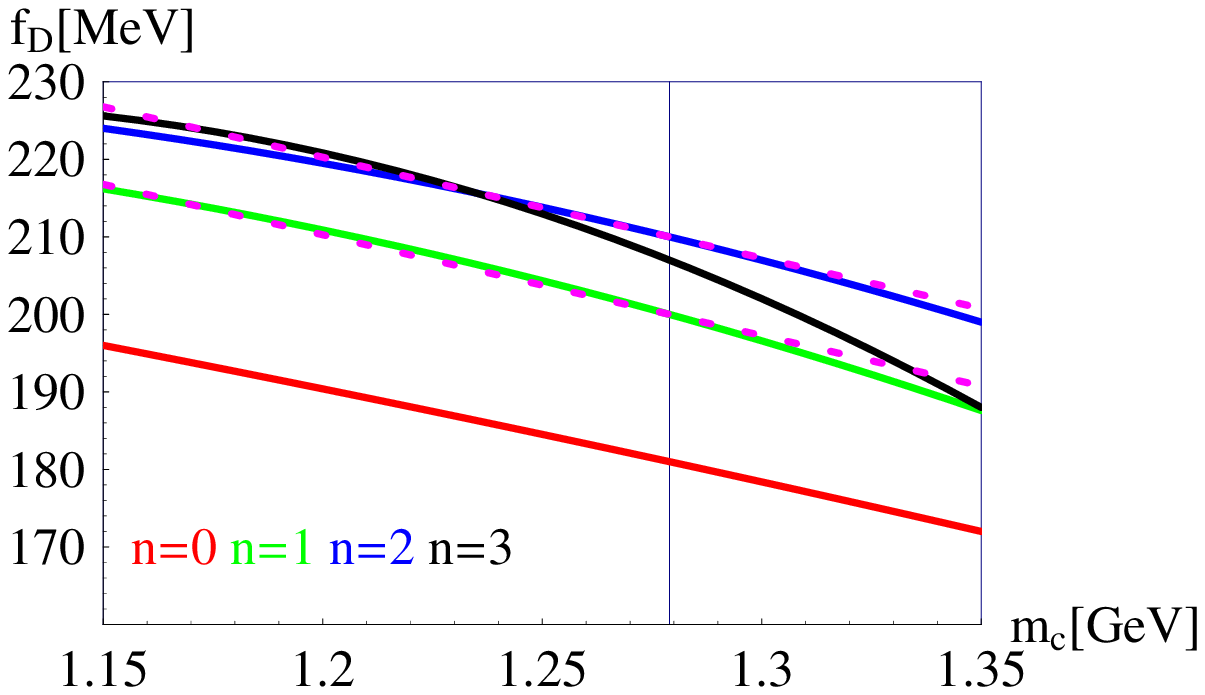}\\
(a) & (b) & (c)
\end{tabular}
\caption{\label{Plot:fD}Dual mass (a) and dual decay constant (b)
of the $D$ meson extracted by adopting different Ans\"atze
(\ref{zeff}) for the~effective continuum threshold $s_{\rm
eff}(\tau)$ and fixing these thresholds according to
(\ref{chisq}). The results for $m_c\equiv {\overline
m}_c({\overline m_c})=1.279$ GeV, $\mu=m_c$, and central values of
the other relevant parameters are presented. (c) Dual decay
constant of the $D$ meson vs.\ $m_c$ for $\mu=m_c$~and central
values of the other OPE parameters. The index $n=0,1,2,3$ denotes
the power of our polynomial Ansatz for the effective continuum
threshold in (\ref{zeff}).}
\end{figure}

The Borel window for the charmed meson is chosen according to the
criteria discussed above: $\tau=(0.1 - 0.5)\;\mbox{GeV}^{-2}$.
Figure~\ref{Plot:fD} illustrates the application of our
prescription of obtaining the effective continuum threshold and
extracting the corresponding $f_D$. We would like to point out
that $\tau$-dependent effective thresholds lead to a much
better~reproduction of the meson mass in the window than a
constant one (Fig.~\ref{Plot:fD}a). This signals that the dual
correlators corresponding~to the $\tau$-dependent thresholds are
less contaminated by the excited states.

The dependence of the extracted value of the $D$-meson decay
constant $f_D$ on the $c$-quark mass $m_c\equiv {\overline
m}_c({\overline m_c})$~and~the quark condensate $\langle \bar
qq\rangle\equiv \langle \bar qq\rangle(2\,{\rm GeV})$ may be
parameterized in the form
\begin{equation}
f_{D}^{\rm dual}(m_c,\mu=m_c,\langle \bar qq\rangle) =\left[206.2
-13\left(\frac{m_c-\mbox{1.279 GeV}}{\mbox{0.1 GeV}}\right) + 4
\left(\frac{|\langle \bar qq\rangle|^{1/3}-\mbox{0.267
GeV}}{\mbox{0.01 GeV}}\right) \pm 5.1_{\rm (syst)} \right]
\mbox{MeV}.
\end{equation}
This relation describes the band of values delimited by the two
dotted lines in Fig.~\ref{Plot:fD}c, which include the
results~derived with the linear, quadratic, and cubic Ans\"atze
for the effective continuum threshold.
\begin{figure}[!ht]
\begin{tabular}{ccc}
\includegraphics[width=5.75cm]{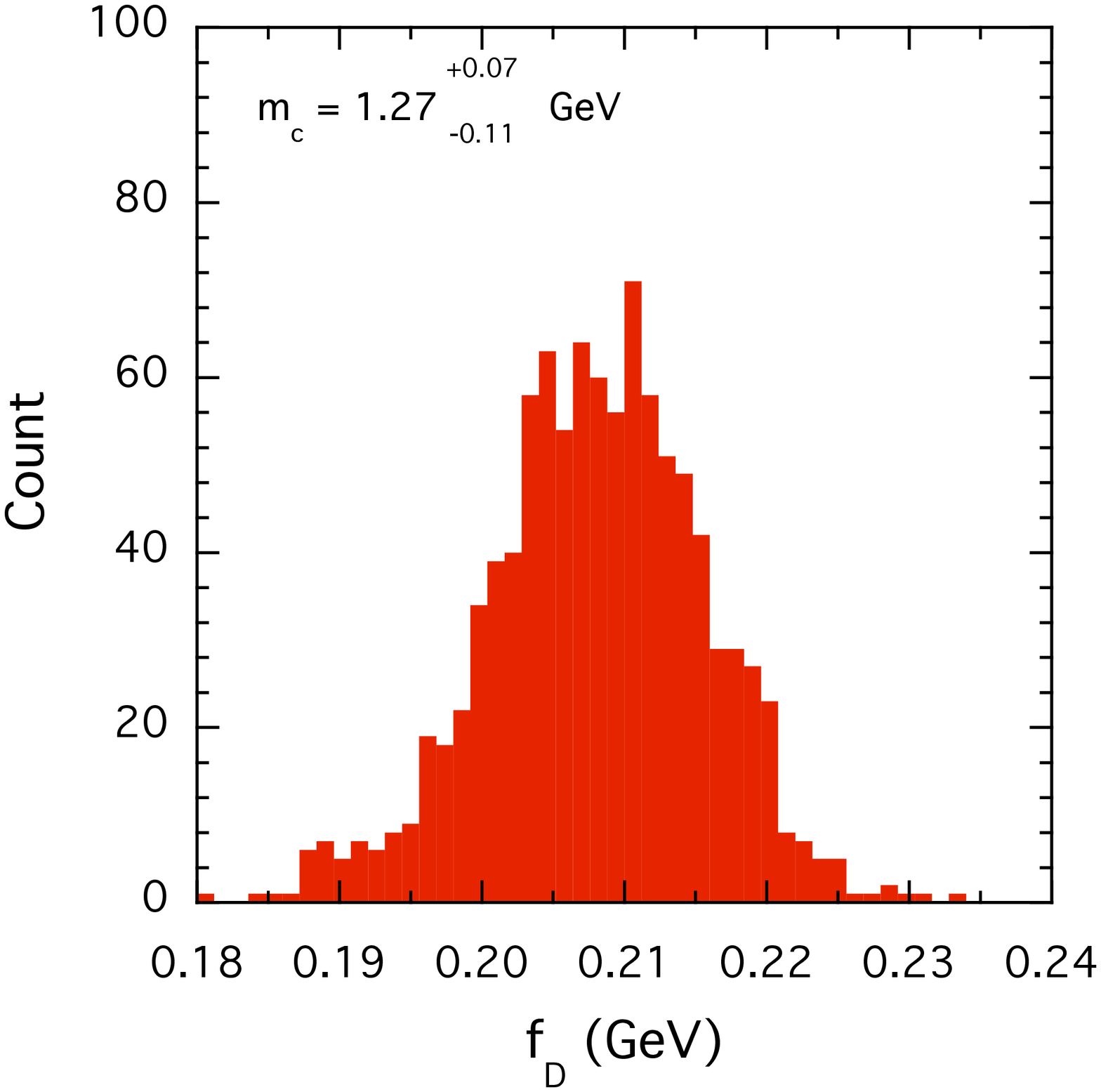}&
\includegraphics[width=5.75cm]{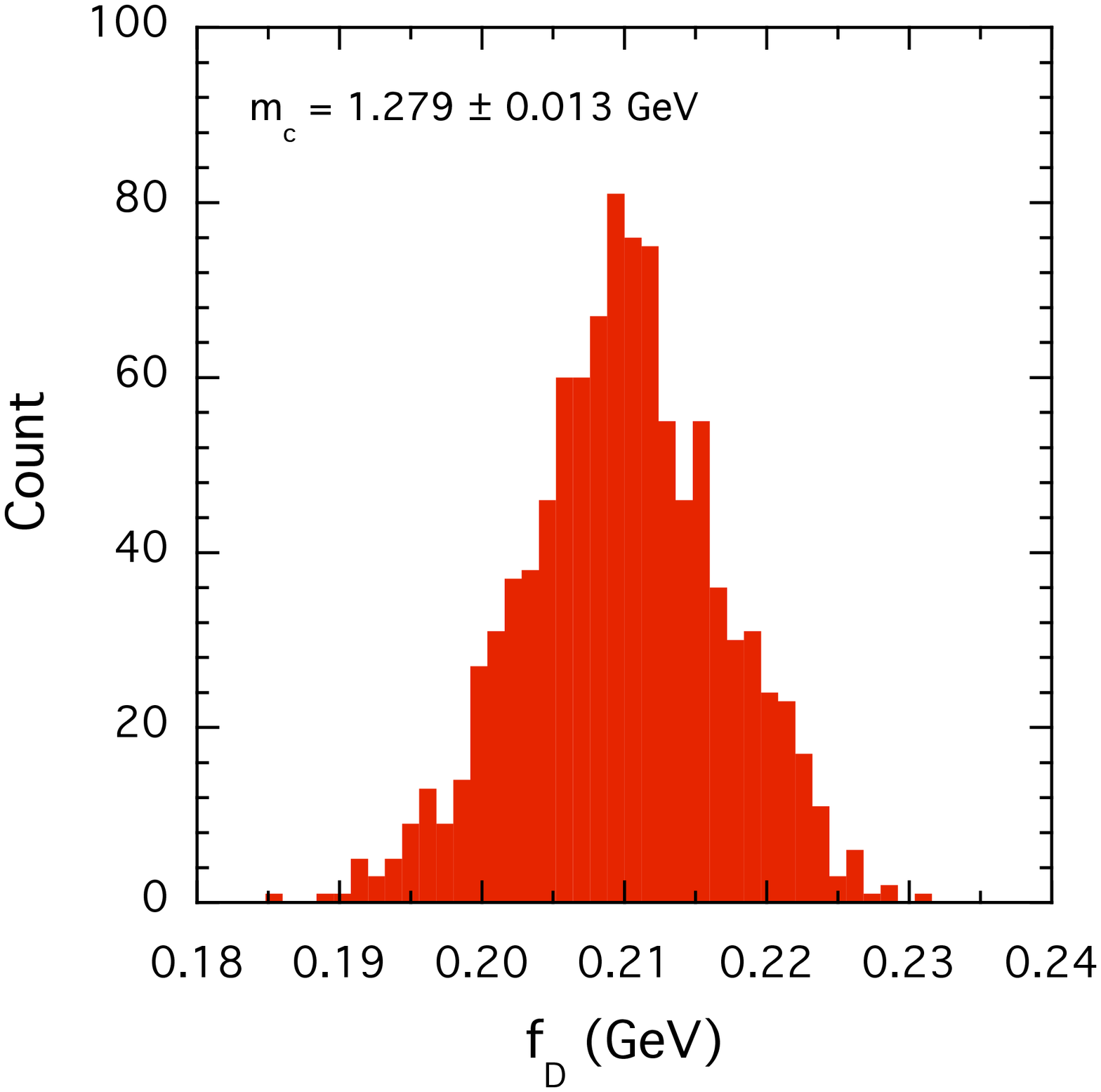}&
\includegraphics[width=5.75cm]{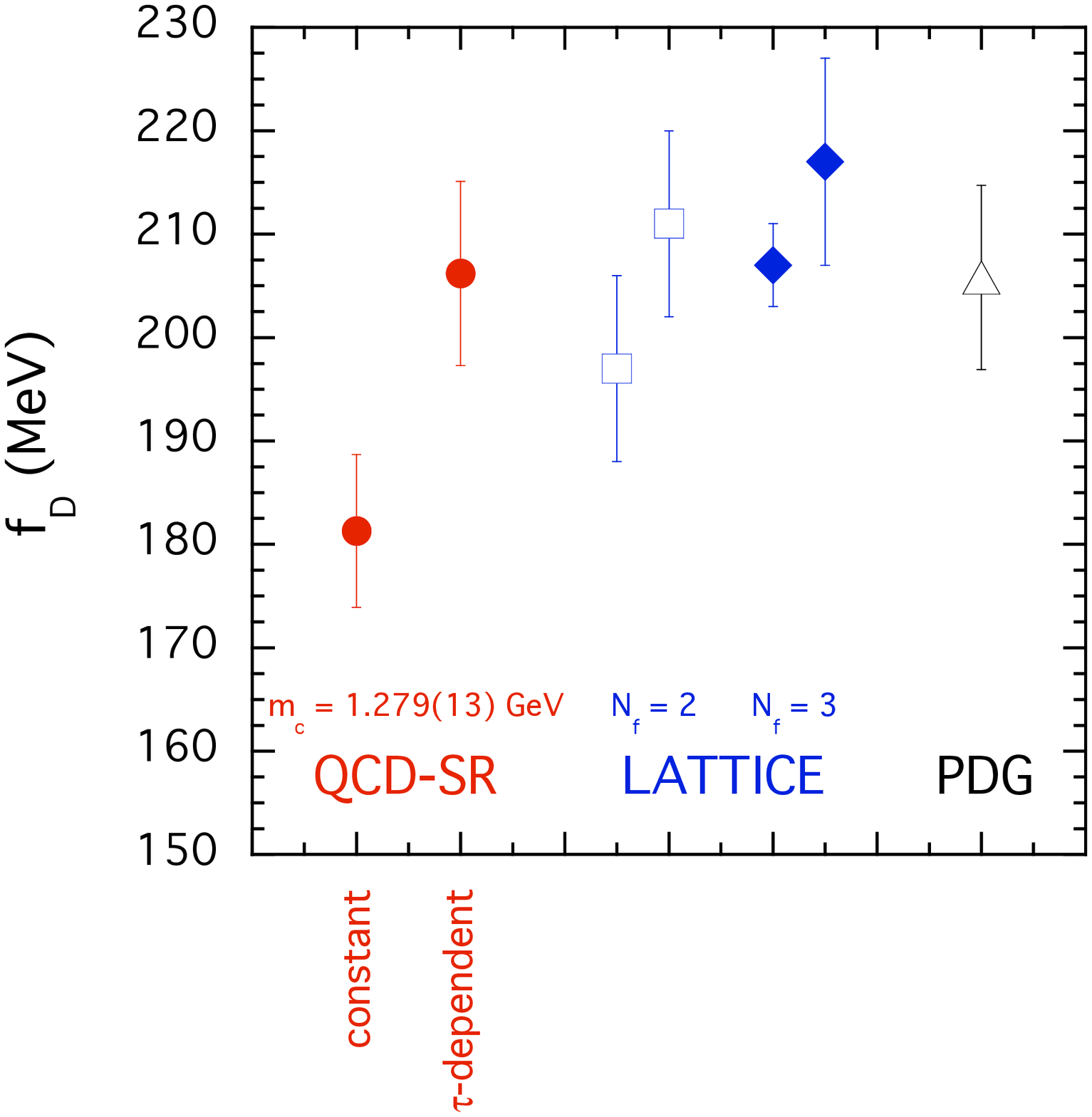}\\
(a) & (b) & (c)
\end{tabular}
\caption{\label{Plot:fD_bootstrap}Distribution of $f_D$ as
obtained by a bootstrap analysis of the OPE uncertainties. For all
OPE parameters but $\mu$ Gaussian distributions with corresponding
errors quoted in (\ref{Eq:OPEP}) are employed. A variation of
$m_c$ in the interval (\ref{Eq:mc1}) (a) and (\ref{Eq:mc2})~(b) is
allowed. For $\mu$ we assume a uniform distribution in the range
$1\;{\rm GeV} < \mu < 3\; {\rm GeV}$. (c) Summary of our results
for $f_D$. For comparison, the lattice results are shown for two
dynamical light flavors ($N_f = 2$) \cite{ETMC1,ETMC2} and three
dynamical~flavors~($N_f = 3$) \cite{HPQCD1,FNAL+MILC1}. The
triangle represents the experimental result from PDG \cite{pdg}.
For the $\tau$-dependent QCD-SR result the depicted~error~is the
sum of the OPE and systematic uncertainties given in (\ref{fD}),
added in quadrature.}
\end{figure}
Figure~\ref{Plot:fD_bootstrap}a displays the results of the
bootstrap analysis of the OPE uncertainties. The distribution has
a Gaussian shape, and therefore the corresponding OPE uncertainty
is the Gaussian error. Adding the width of the band provided by
the $\tau$-dependent $n=1,2,3$ Ans\"atze for the effective
continuum threshold as the (intrinsic) systematic error of our
approach, we obtain the following result:
\begin{equation}
\label{fD} f_{D} = (206.2 \pm 7.3_{\rm (OPE)} \pm 5.1_{\rm
(syst)})\; \mbox{MeV}.
\end{equation}
We have considered for $m_c$ the two ranges in (\ref{Eq:mc1}) and
(\ref{Eq:mc2}). The OPE-related error is practically the same
for~both ranges (see Figs.~\ref{Plot:fD_bootstrap}a,b), so the
main source of the OPE uncertainty in the extracted $f_D$ comes
from the OPE parameters other than $m_c$ (mainly, the
renormalization scale and the quark condensate).

Notice that the bootstrap procedure for a $\tau$-independent
effective threshold gives a substantially lower $f_D$ range,~viz.,
%\begin{eqnarray}
%\label{fD_constant}
$f_D{(n=0)} = (181.3\pm 7.4_{\rm (OPE)})\; \mbox{MeV}$,
%\end{eqnarray}
which deviates from our $\tau$-dependent result (\ref{fD}) by
almost three times the~OPE uncertainty. Moreover, as we have
already shown in our previous works \cite{lms_2ptsr}, making use
of merely a constant Ansatz~for the effective continuum threshold
does not allow one to probe at all the intrinsic systematic error
of the QCD sum~rule. From (\ref{fD}) the latter turns out to be of
the same order as the OPE uncertainty.

The $\tau$-dependent threshold leads to a clearly discernible
effect and brings the results from QCD sum rules into~perfect
agreement with recent lattice results and the data
(Fig.~\ref{Plot:fD_bootstrap}b). A perfect agreement of our result
with the lattice ones~and with experiment provides a further
confirmation of the reliability of our procedure.

\subsection{\boldmath Decay constant of the $D_s$ meson}
\begin{figure}[!ht]
\begin{tabular}{ccc}
\includegraphics[width=5.75cm]{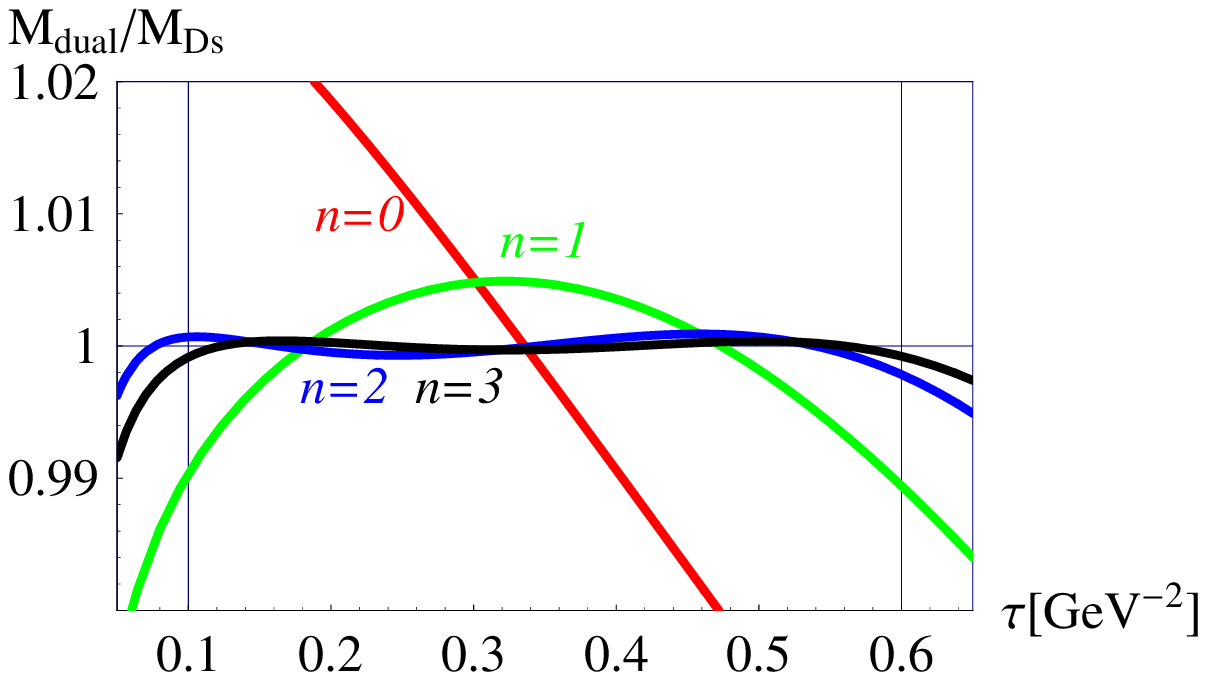}&
\includegraphics[width=5.75cm]{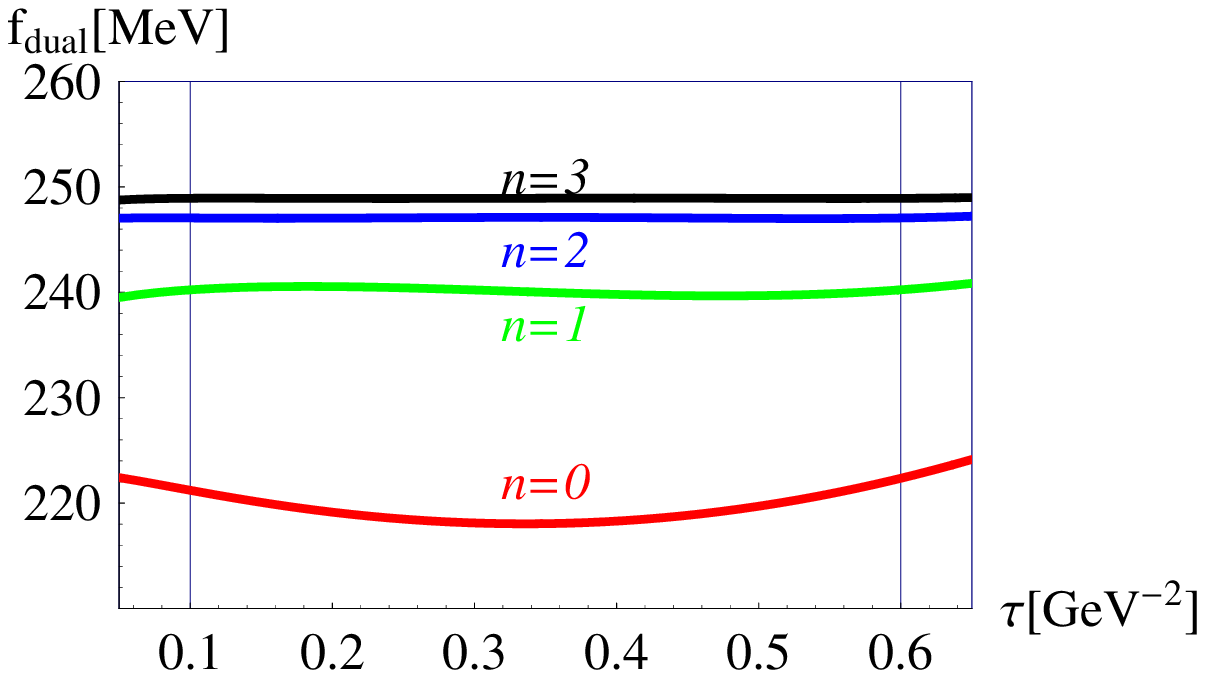}&
\includegraphics[width=5.75cm]{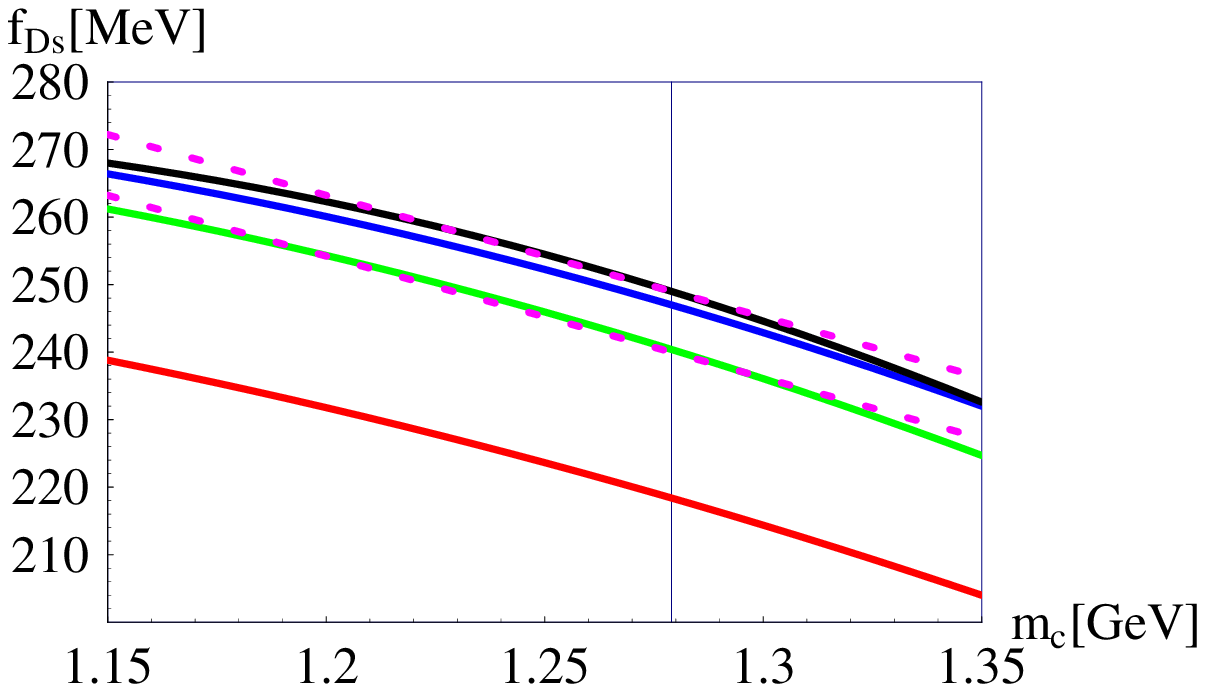}\\
(a) & (b) & (c)
\end{tabular}
\caption{\label{Plot:fDs}Same as Fig.~\ref{Plot:fD} but for the
$D_S$ meson.}
\end{figure}

The corresponding $\tau$-window is $\tau =
(0.1-0.6)\;\mbox{GeV}^{-2}$. Figure~\ref{Plot:fDs} provides the
details of our extraction procedure.~Our
results for $f_{D_s}$ may be represented in the form
\begin{equation}
f_{D_s}^{\rm dual}(m_c,\mu=m_c,\langle \bar ss\rangle)
=\left[245.3 -18\left(\frac{m_c-\mbox{1.279 GeV}}{\mbox{0.1
GeV}}\right) + 3.5 \left(\frac{|\langle \bar
ss\rangle|^{1/3}-\mbox{0.248 GeV}}{\mbox{0.01 GeV}}\right) \pm
4.5_{\rm (syst)} \right] \mbox{MeV}.
\end{equation}
This relation describes the band of $f_{D_s}$ values as a function
of $m_c\equiv {\overline m}_c({\overline m_c)}$ indicated by the
two dotted lines~in~Fig.~\ref{Plot:fDs}c, as well as the
dependence on the quark condensate $\langle \bar ss\rangle\equiv
\langle \bar ss\rangle(2\,{\rm GeV})$.
\begin{figure}[b]
\begin{tabular}{ccc}
\includegraphics[width=6.5cm]{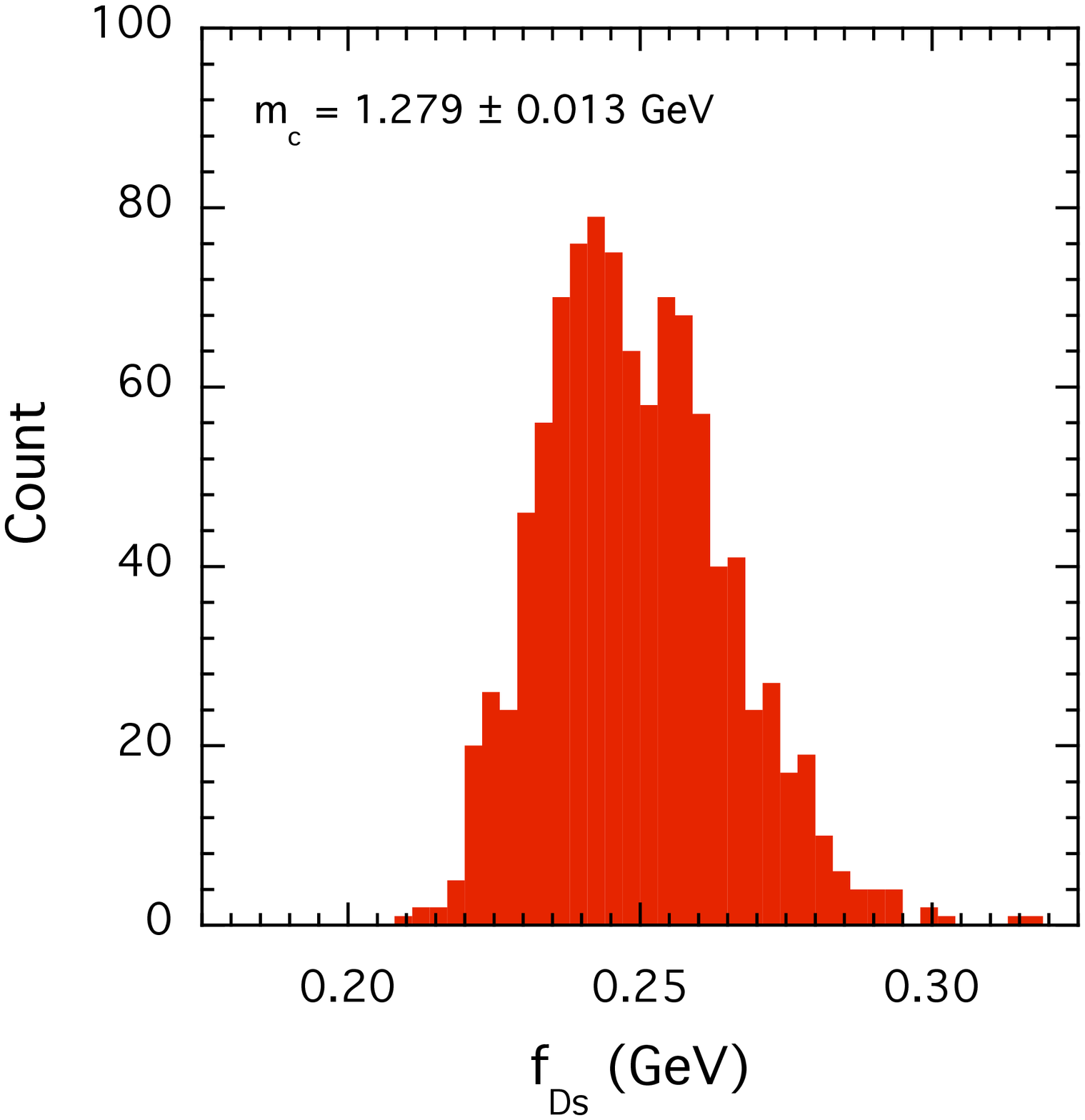}&\qquad \qquad&
\includegraphics[width=6.5cm]{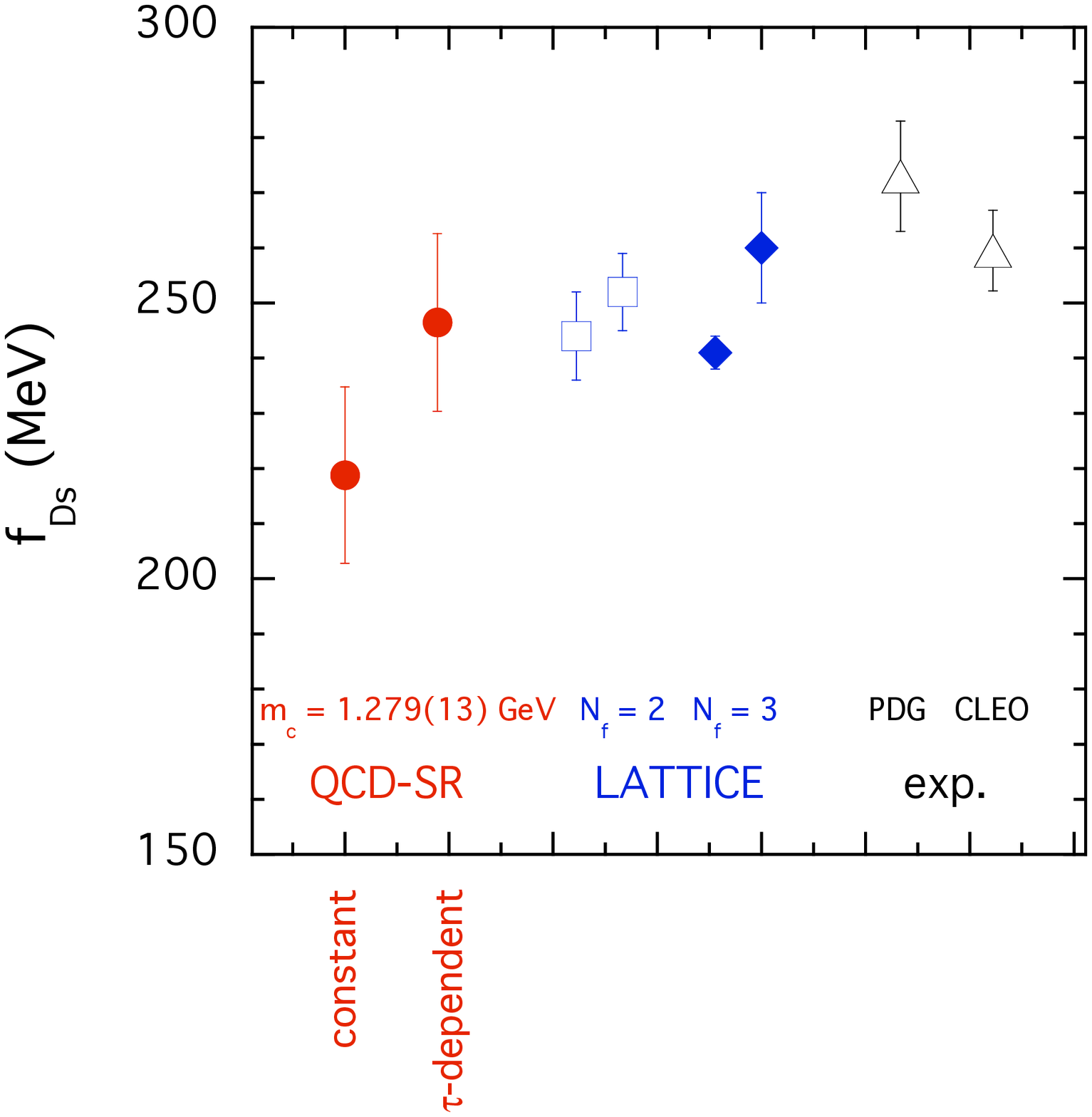}
\\
(a) &\qquad \qquad& (b)
\end{tabular}
\caption{\label{Plot:fDs_bootstrap}(a) Distribution of $f_{Ds}$ as
obtained by the bootstrap analysis of the OPE uncertainties. For
all our OPE parameters~but $\mu$ Gaussian distributions with
corresponding errors given in (\ref{Eq:OPEP}) are adopted. A
variation of $m_c$ in the interval (\ref{Eq:mc2}) is~allowed. For
$\mu$ we assume a uniform distribution in the range $1\;{\rm GeV}
< \mu < 3\; {\rm GeV}$. (b) Summary of our results for $f_{Ds}$.
Lattice results are depicted for $N_f=2$ \cite{ETMC1,ETMC2} and
$N_f=3$ \cite{HPQCD1,FNAL+MILC1}. The experimental results are
from PDG \cite{pdg} and the CLEO~collaboration~\cite{cleo}. For
the $\tau$-dependent QCD-SR result the error shown is the sum of
the OPE and systematic uncertainties given in (\ref{fDs}),
added~in quadrature.}
\end{figure}
Performing the bootstrap~analysis of the OPE uncertainties, we
obtain the following estimate:
\begin{equation}
\label{fDs} f_{D_s} = (245.3 \pm 15.7_{\rm (OPE)} \pm 4.5_{\rm
(syst)})\; {\rm MeV}.
\end{equation}
As in the case of $f_D$, a constant threshold yields a
substantially lower $f_D$ value: $f_{D_s}{(n=0)} = (218.8 \pm
16.1_{\rm (OPE)})\;{\rm MeV}$.

\subsection{\boldmath $f_{D_s}/f_{D}$}
For the ratio of the $D_{(s)}$-meson decay constants, we find
\begin{eqnarray}
\label{ratioD} f_{D_s}/f_D = 1.193\pm 0.025_{(\rm OPE)}\pm
0.007_{(\rm syst)},
\end{eqnarray}
to be compared with the PDG average, $f_{D_s}/f_D = 1.25\pm 0.06$
\cite{pdg}, and the recent lattice results $f_{D_s}/f_D = 1.24 \pm
0.03$~\cite{ETMC1} at $N_f=2$, and $f_{D_s}/f_D = 1.164 \pm 0.011$
\cite{HPQCD1} and $f_{D_s}/f_D = 1.20 \pm 0.02$ \cite{FNAL+MILC1}
at $N_f=3$. The error in (\ref{ratioD}) comes~mainly from the
uncertainties in the quark condensates: $\langle \bar s
s\rangle/\langle \bar q q\rangle=0.8\pm 0.3$.

\section{\boldmath Decay constants of the $B$ and $B_s$ mesons}\label{Sec:BDC}
We set the Borel window as
%\begin{eqnarray}
$\tau = (0.05 - 0.18)\; {\rm GeV}^{-2}$.
%\end{eqnarray}
%In the region $\tau < 0.18$ GeV$^{-2}$ power corrections give less than 30\%
%of the cut perturbative contribution to the correlator.
Note that the radiative corrections to the condensates~increase
rather fast with $\tau$, so it is preferable to stay at lower
values of $\tau$.

\subsection{\boldmath Decay constant of the $B$ meson}
\begin{figure}[!hb]
\begin{tabular}{ccc}
\includegraphics[width=5.75cm]{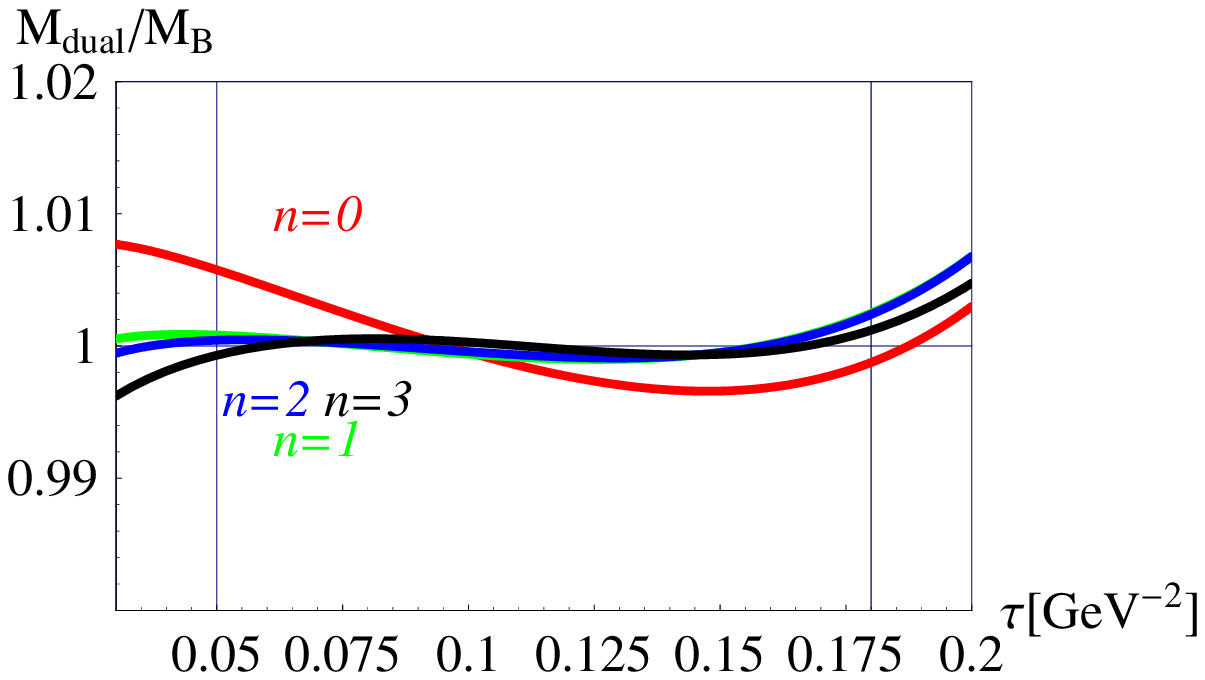}&
\includegraphics[width=5.75cm]{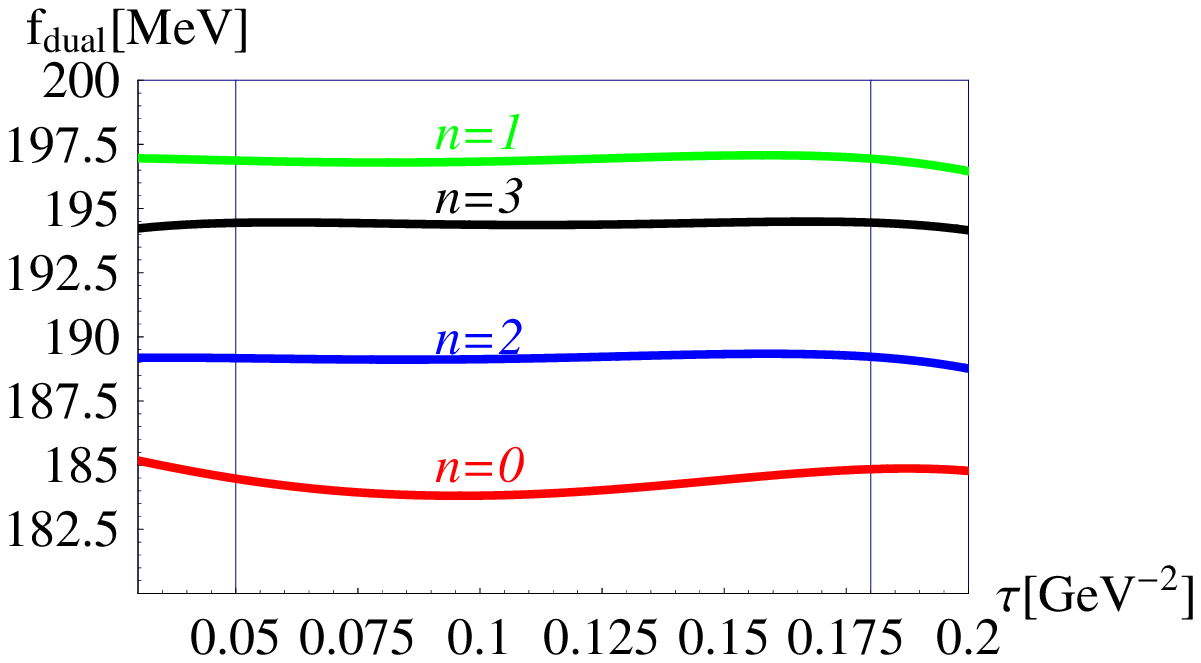}&
\includegraphics[width=5.75cm]{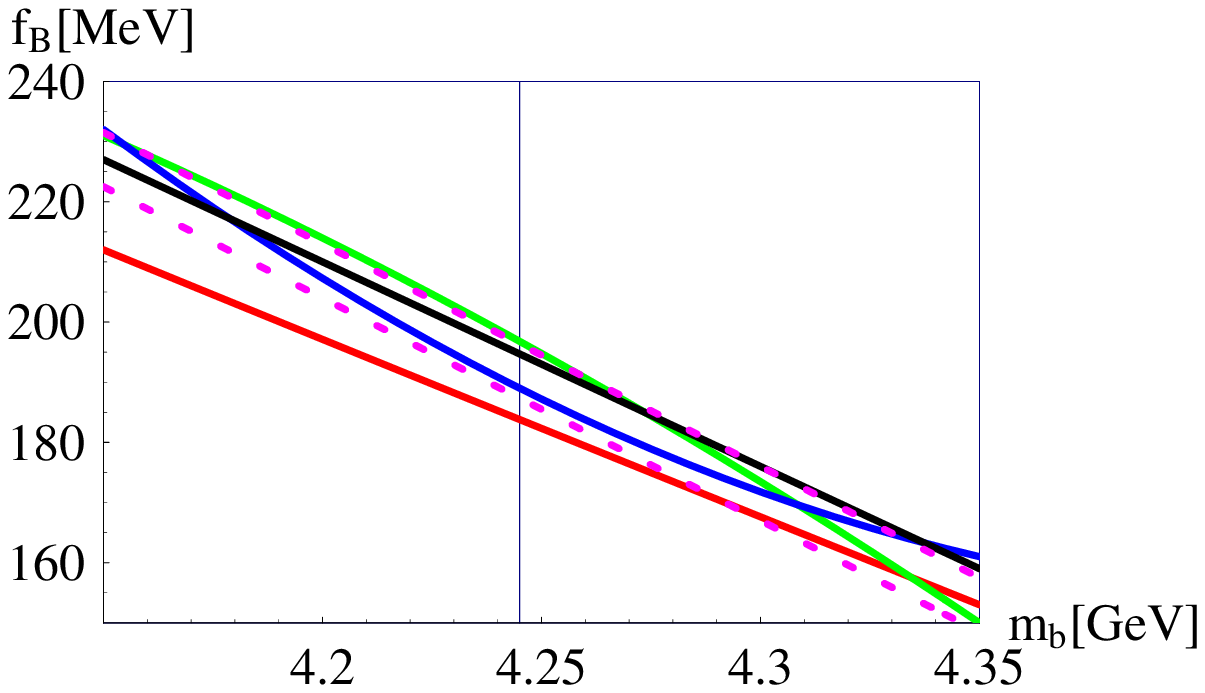}\\
(a) & (b) & (c)
\end{tabular}
\caption{\label{Plot:fB}Dual mass (a) and dual decay constant (b)
of the $B$ meson obtained by using different Ans\"atze for the
effective continuum threshold $s_{\rm eff}(\tau)$ (\ref{zeff}) and
fixing the coefficients according to (\ref{chisq}). The results
for $m_b\equiv {\overline m}_b({\overline
m_b})=4.245\;\mbox{GeV}$,~$\mu=m_b$,~and central values of all
other relevant parameters are presented. (c) Dual decay constant
of the $B$ meson vs.\ $m_b$ for $\mu=m_b$ and central values of
all other OPE parameters. The index $n=0,1,2,3$ denotes the power
of the polynomial Ansatz for the effective continuum threshold in
(\ref{zeff}).}
\end{figure}

Figure~\ref{Plot:fB} shows the application of our prescription to
the extraction of $f_B$. The correlator $\Pi(\tau)$, which has
dimension six, is extremely sensitive to the precise value of
$m_b$. The dependence of the extracted value of the decay
constant~$f_B$~on the $b$-quark mass $m_b\equiv {\overline
m}_b({\overline m_b})$ and the condensate $\langle \bar
qq\rangle\equiv \langle \bar qq\rangle(2\,{\rm GeV})$ may be
parameterized in the form
\begin{eqnarray}
f_{B}^{\rm dual}(m_b,\mu=m_b,\langle \bar qq\rangle) =\left[193.4
- 37\left(\frac{m_b-\mbox{4.245 GeV}}{\mbox{0.1 GeV}}\right) + 4
\left(\frac{|\langle \bar qq\rangle|^{1/3}-\mbox{0.267
GeV}}{\mbox{0.01 GeV}}\right) \pm 4_{\rm (syst)} \right]
\mbox{MeV}.
\end{eqnarray}
The above relation describes the band of values delimited by the
two dotted lines in Fig.~\ref{Plot:fB}c which, as before,
include~the results obtained with the linear, quadratic, and cubic
Ans\"atze for the effective continuum threshold. In addition,
it~also encodes the dependence on the value of the quark
condensate.

We now perform a bootstrap analysis of $f_B$ combining all OPE
uncertainties. We assume Gaussian distributions~for the OPE
parameters (quark masses, condensates) with corresponding errors.
The renormalization scale $\mu$ is assumed~to be uniformly
distributed in the interval $2\le \mu\; (\mbox{GeV}) \le 8$.

\begin{figure}[!ht]
\begin{tabular}{ccc}
\includegraphics[width=5.5cm]{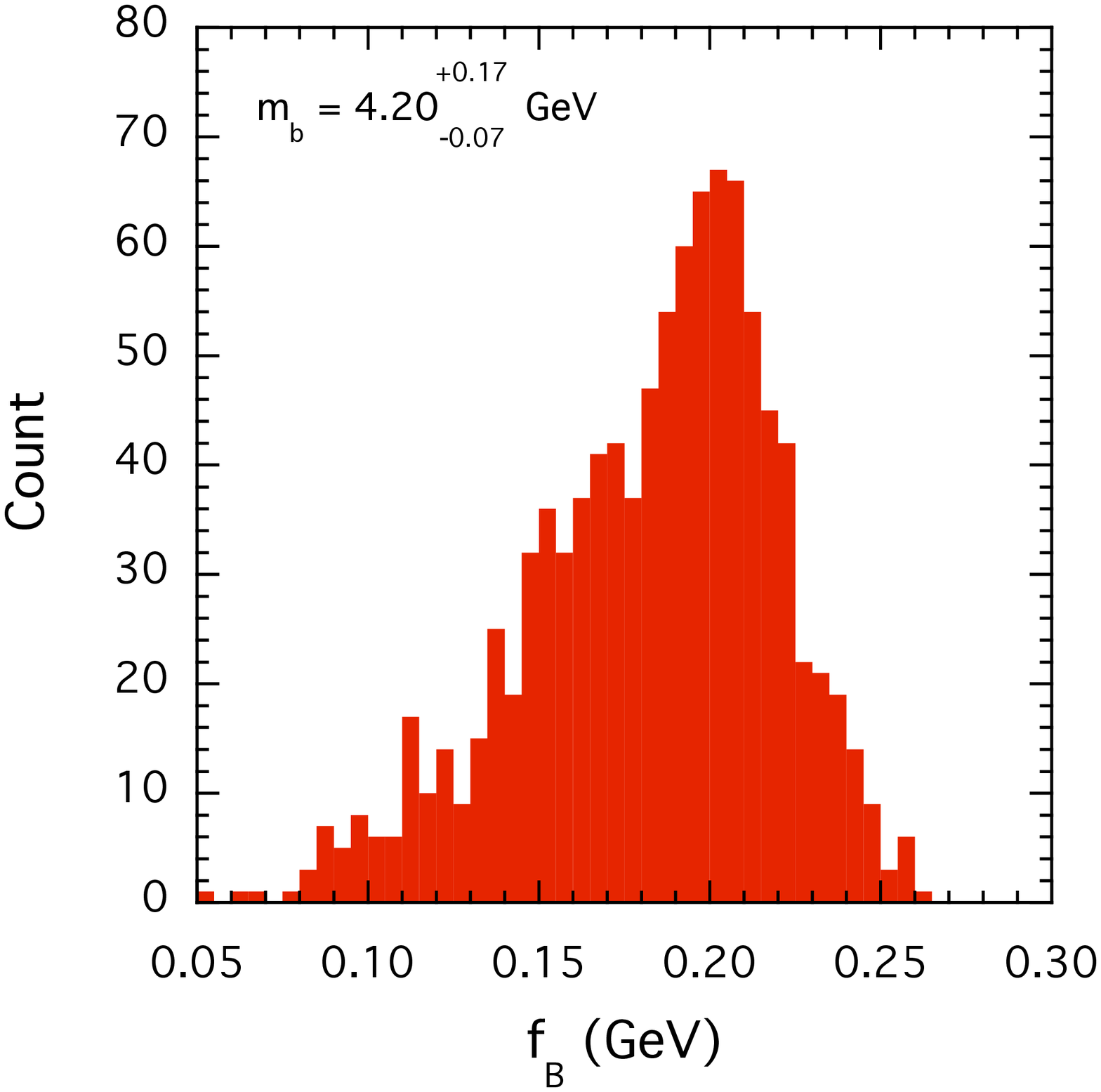}&
\includegraphics[width=5.5cm]{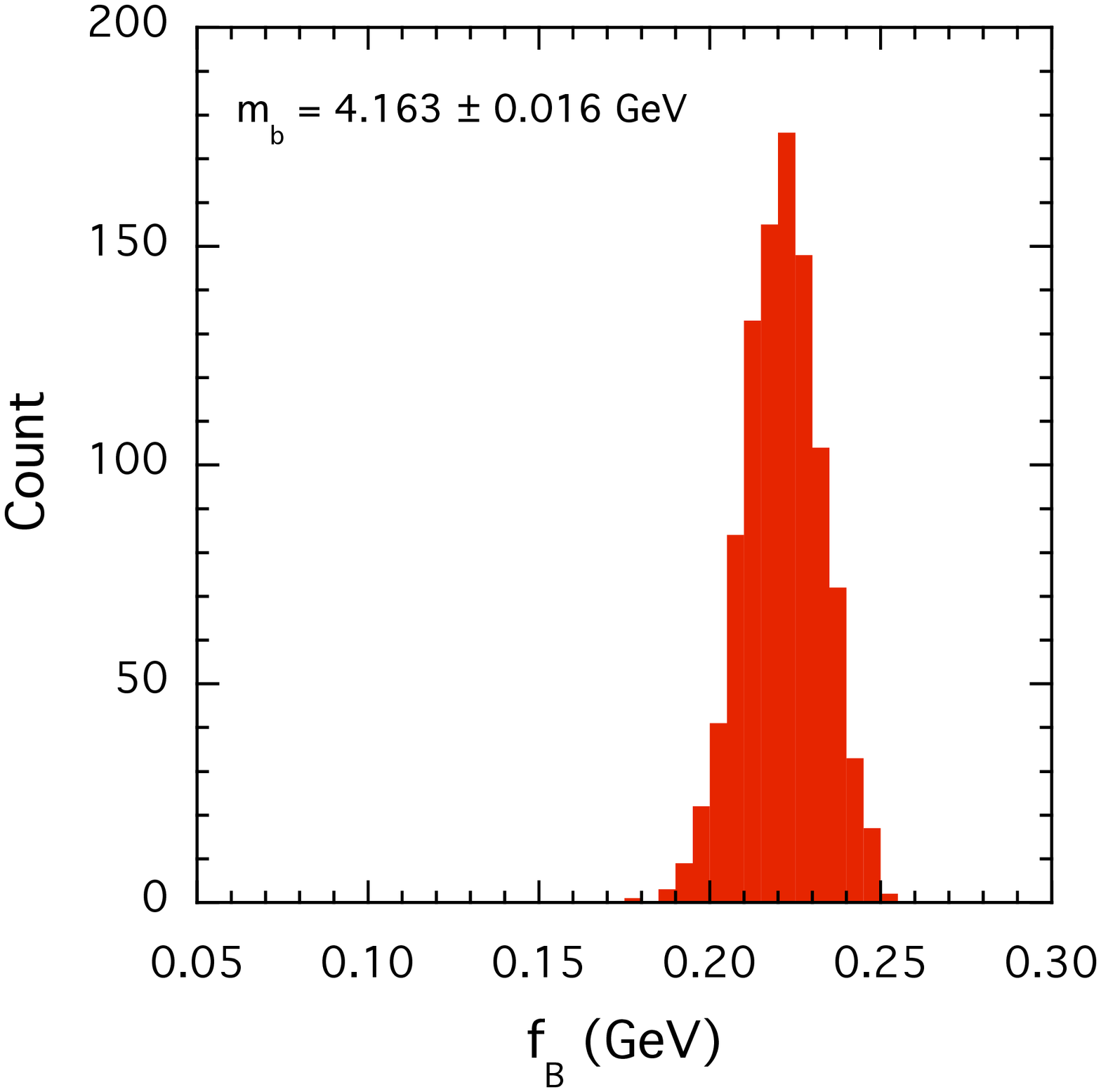}&
\includegraphics[width=5.5cm]{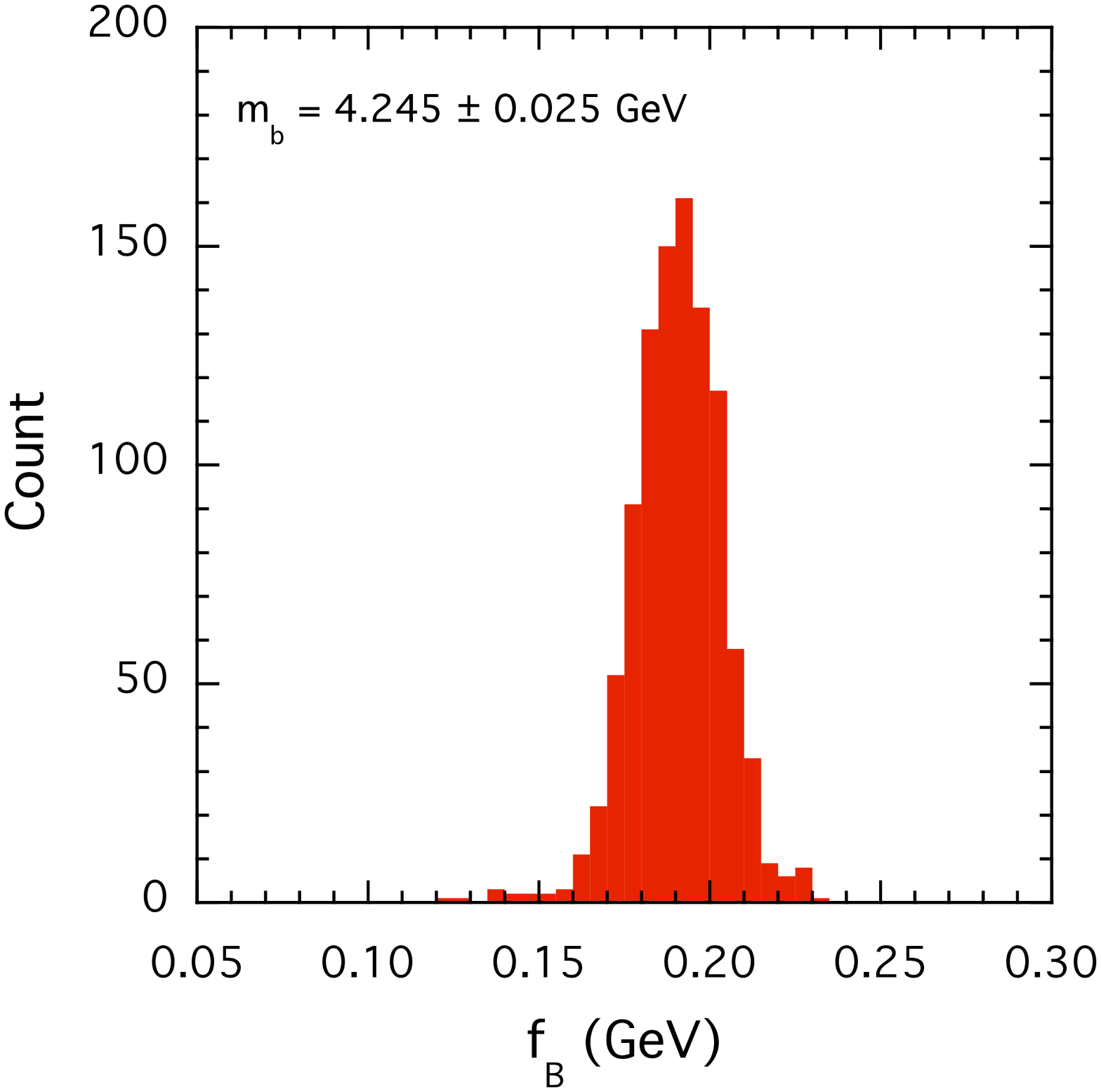}\\
(a) & (b) & (c)
\end{tabular}
\caption{\label{Plot:fB_bootstrap}Distribution of $f_B$ for three
ranges of $m_b$: the range (\ref{Eq:mc1}) (a), the range
(\ref{Eq:mc2}) (b), and the range specified by (\ref{mb})~(c).}
\end{figure}

Because of the high sensitivity of the correlator to the $b$-quark
mass, the sum-rule estimate for $f_B$ strongly depends on the
range of $m_b$ used. For the PDG range
$m_b=\left(4.2^{+0.17}_{-0.07}\right)\mbox{GeV}$ \cite{pdg}, the
OPE uncertainties are very large;~therefore,~no reasonable
estimate of $f_B$ may be obtained (see
Fig.~\ref{Plot:fB_bootstrap}).
\begin{figure}[!ht]
\begin{tabular}{c}
\includegraphics[width=8cm]{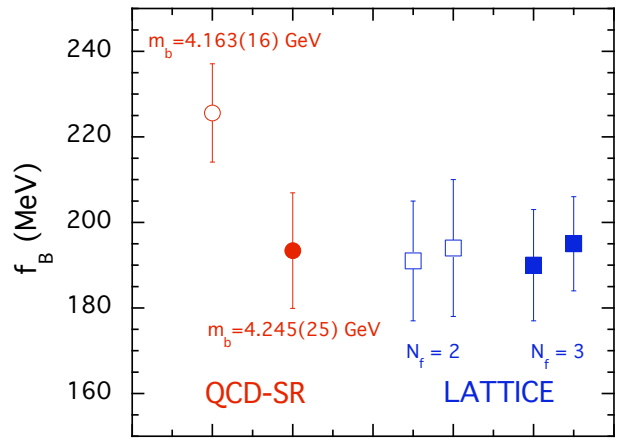}
\end{tabular}
\caption{\label{Plot:fB_summary}Our estimates for $f_B$, with
$m_b$ in the ranges (\ref{Eq:mc2}) and (\ref{mb}), vs.\ recent
results from lattice QCD for $N_f=2$ \cite{ETMC2,ETMC3}~and
$N_f=3$ \cite{HPQCD2,FNAL+MILC2}.}
\end{figure}

Adopting the recently reported precise range $m_b=(4.163\pm
0.016)\;\mbox{GeV}$ from \cite{mb} leads to a rather accurate
estimate:
%\begin{eqnarray}
%\label{fB}
$f_{B} = (225.6 \pm 11.3_{\rm (OPE)} \pm 2.2_{\rm (syst)})\; {\rm
MeV}$.
%\end{eqnarray}
However, we observe (see Fig.~\ref{Plot:fB_summary}) some tension
between this value and recent lattice calculations of $f_B$, which
yield, on average, $f_B{\rm (lattice)}=(193\pm 130)\;\mbox{MeV}$.
Requiring the sum-rule~estimate~to match the lattice average leads
to a rather accurate determination of $m_b$:
\begin{eqnarray}
\label{mb} {\overline m}_b({\overline m_b}) = (4.245\pm 0.025)\;
{\rm GeV}.
\end{eqnarray}
This, however, differs considerably from the range found in
\cite{mb} as well as  from the recent finding 
${\overline m}_b({\overline m_b}) = 4.164 \pm 0.023\; 
{\rm GeV}$  \cite{mb_HPQCD}, obtained from a (perturbative) 
QCD analysis similar to the one used in \cite{mb} but applied 
to the moments of heavy-quark current-current correlators 
calculated in lattice QCD with $N_f = 3$ .
The value (\ref{mb}) is in good agreement with the 
lattice determinations ${\overline m}_b({\overline m_b}) = 
4.26 \pm 0.03_{\mbox{stat}} \pm 0.09_{\mbox{syst}}\; {\rm GeV}$ 
\cite{mb_Gimenez} and ${\overline m}_b({\overline m_b}) = 
4.25 \pm 0.02_{\mbox{stat}} \pm 0.11_{\mbox{syst}}\; {\rm GeV}$ 
\cite{mb_UKQCD}, as well as with the recent preliminary 
result of the Alpha Collaboration \cite{garron}, all of them 
obtained using HQET on the lattice with $N_f=2$. 

The B-meson decay constant and its uncertainties corresponding 
to the $b$-quark mass (\ref{mb}) are
\begin{eqnarray}
\label{fB} f_{B} = (193.4 \pm 12.3_{\rm (OPE)} \pm 4.3_{\rm (syst)})\; {\rm MeV}.
\end{eqnarray}
Finally, let us mention that, for the range (\ref{mb}), a constant
effective threshold gives $f_{B}{(n=0)} = (184 \pm 13_{\rm
(OPE)})\; {\rm MeV}$.

\subsection{\boldmath Decay constant of the $B_s$ meson}
\begin{figure}[!hb]
\begin{tabular}{ccc}
\includegraphics[width=5.5cm]{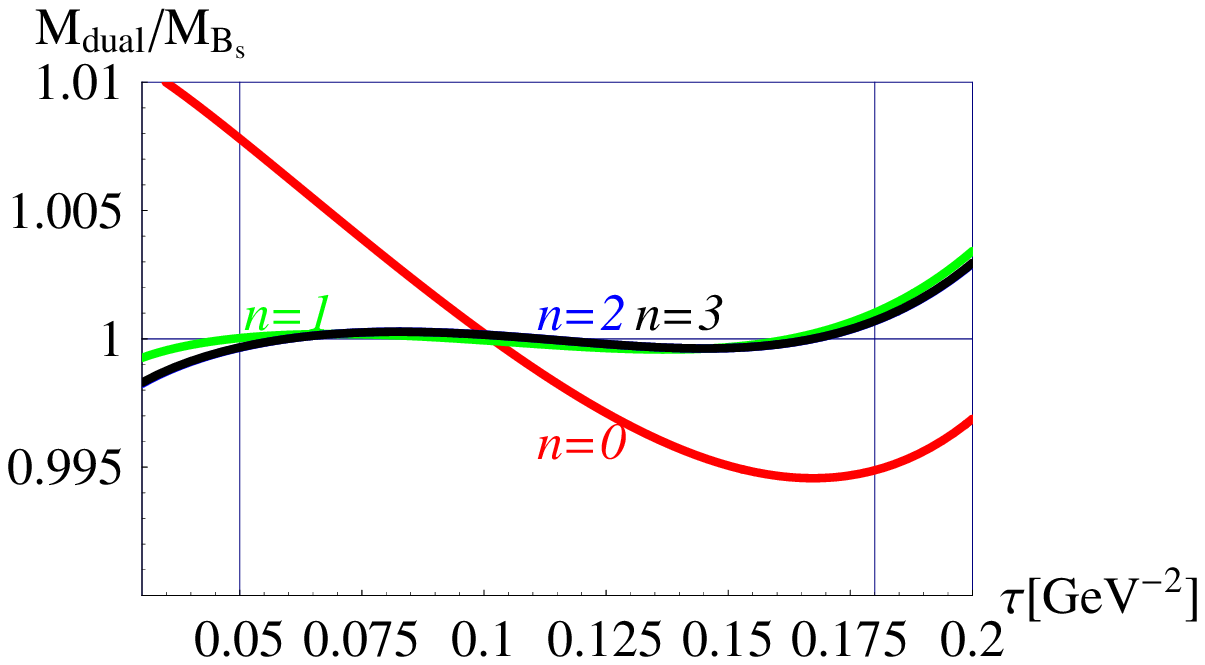}&
\includegraphics[width=5.5cm]{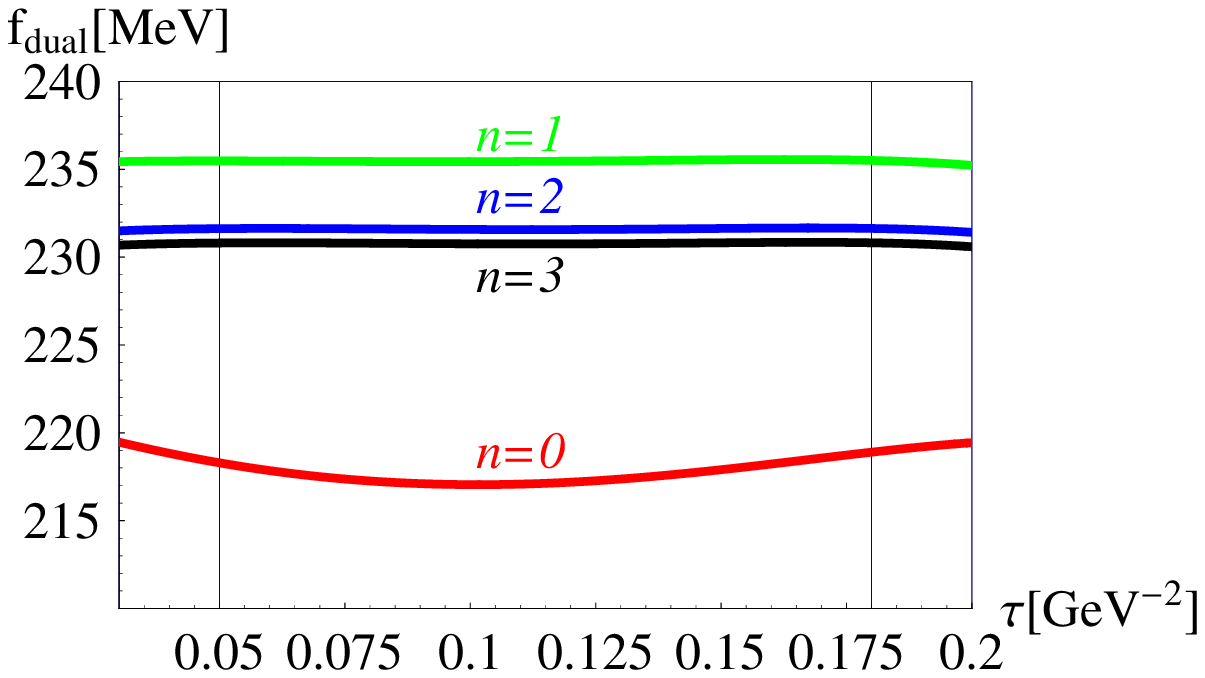}&
\includegraphics[width=5.5cm]{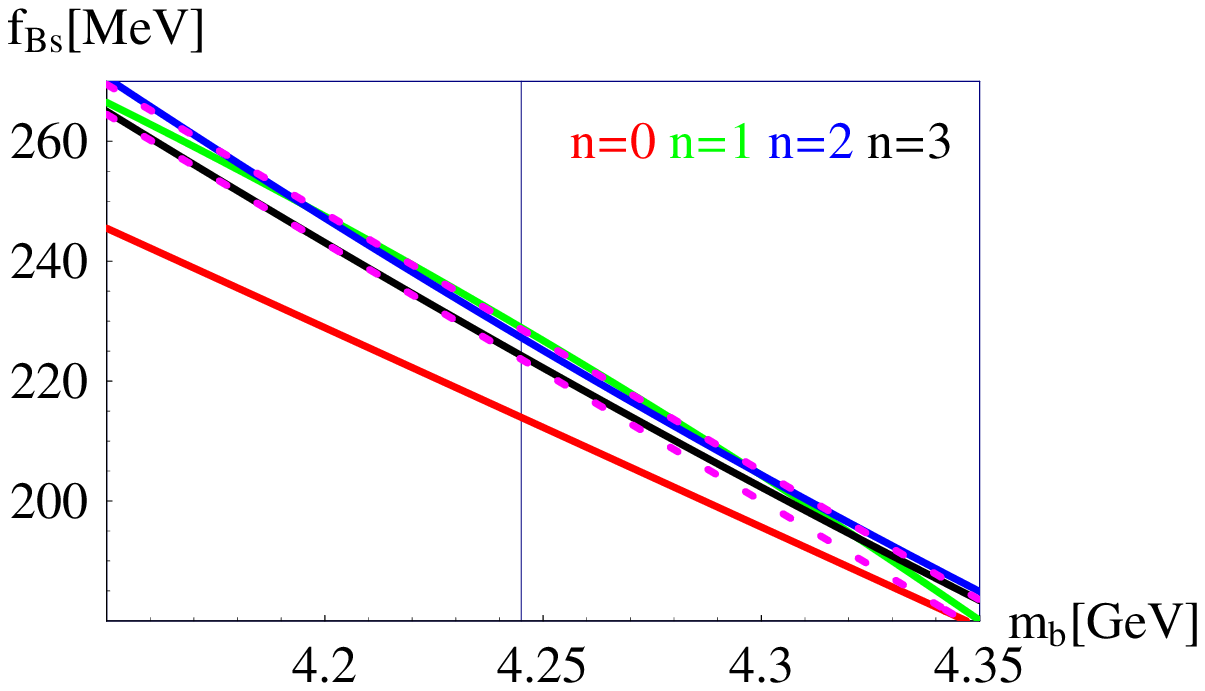}\\
(a) & (b) & (c)
\end{tabular}
\caption{\label{Plot:fBs}Extraction of $f_{B_s}$. Line
identification as in Fig.~\ref{Plot:fB}.}
\end{figure}

Figure~\ref{Plot:fBs} depicts the application of our procedure to
the extraction of $f_{B_s}$.
The dependence of the extracted value~of the decay constant
$f_{B_s}$ on the $b$-quark mass and the strange-quark condensate
is given by the relation
\begin{eqnarray}
f_{B_s}^{\rm dual}(m_b,\mu=m_b,\langle \bar ss\rangle)
=\left[232.5 -43\left(\frac{m_b-\mbox{4.245 GeV}}{\mbox{0.1
GeV}}\right) + 3.5 \left(\frac{|\langle \bar
ss\rangle|^{1/3}-\mbox{0.248 GeV}}{\mbox{0.01 GeV}}\right) \pm
2.4_{\rm (syst)} \right] \mbox{MeV}.
\end{eqnarray}
This formula describes the band of values indicated by two dotted
lines in Fig.~\ref{Plot:fBs}c and, in addition, gives the
dependence on the value of the quark condensate at renormalization
scale $\mu=2\;\mbox{GeV}$.

\begin{figure}[!ht]
\begin{tabular}{ccc}
\includegraphics[width=7.cm]{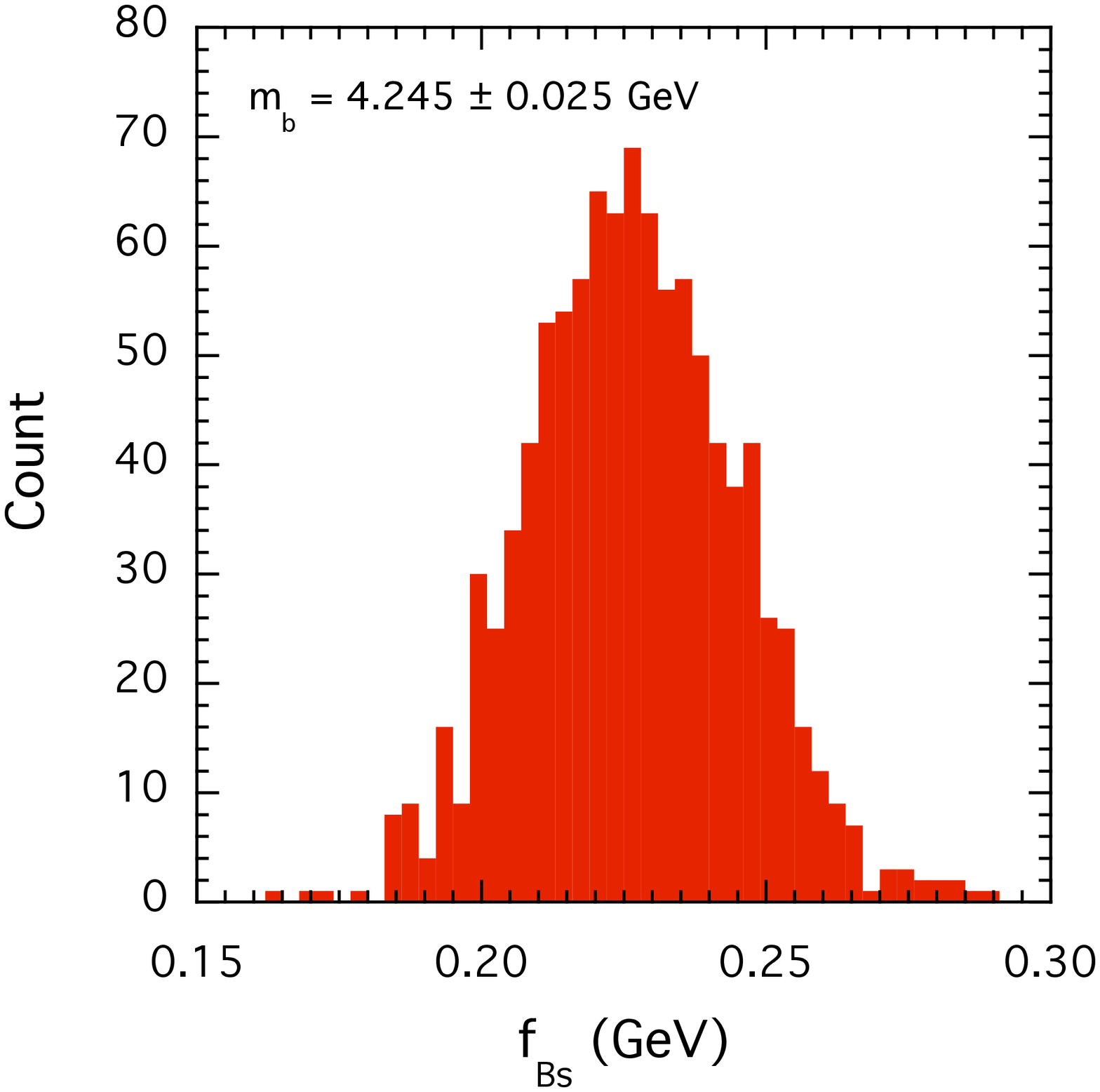}&\qquad \qquad&
\includegraphics[width=7.cm]{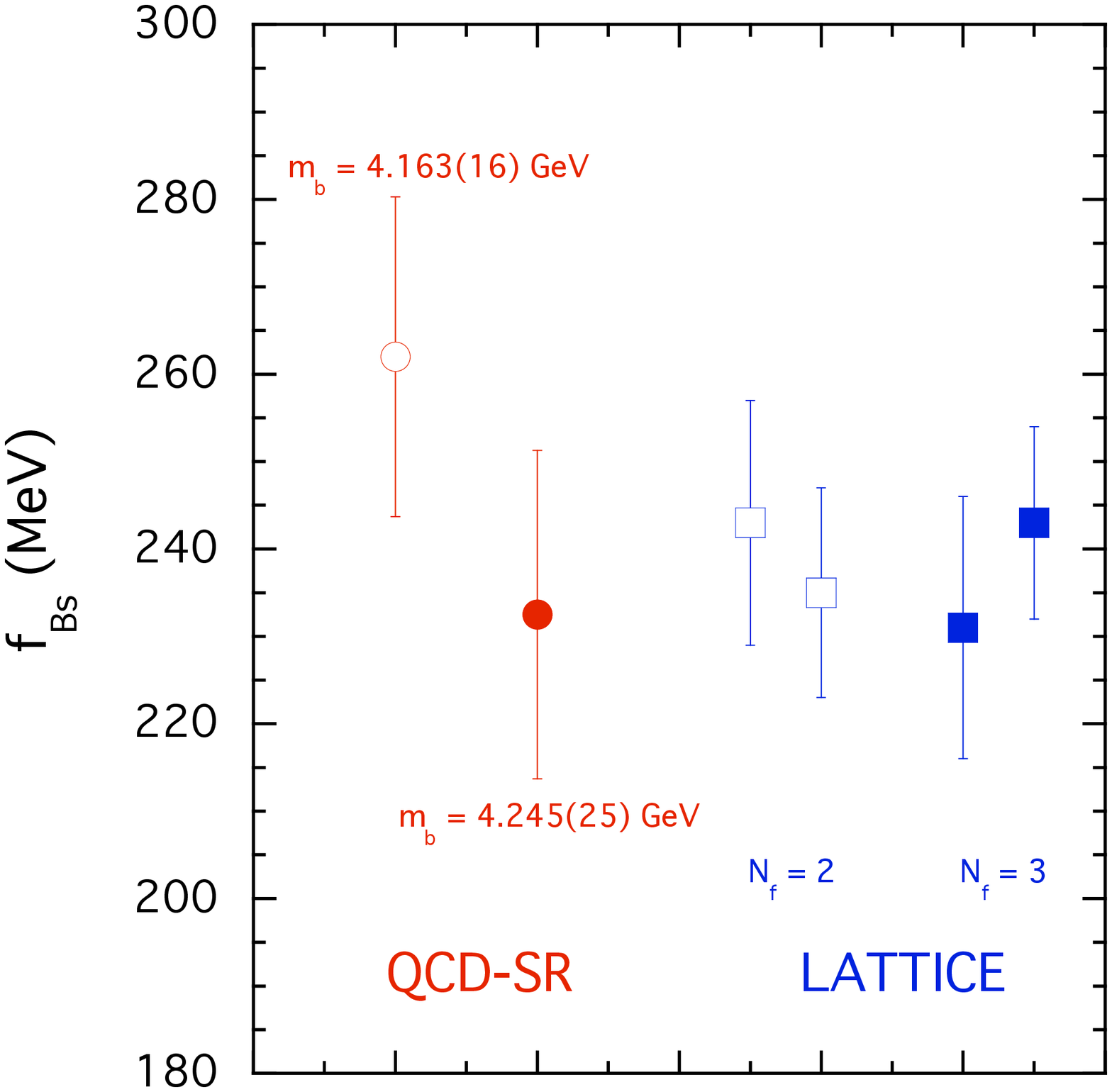}\\
(a) & \qquad \qquad & (b)
\end{tabular}
\caption{\label{Plot:fBs_bootstrap}(a) Distribution of $f_{B_s}$
for $m_b=(4.245\pm 0.025)\;\mbox{GeV}$ [see (\ref{mb})]. (b) Our
results, corresponding to the $m_b$-ranges (\ref{Eq:mc2}) and
(\ref{mb}), vs.\ lattice-QCD results for $N_f=2$
\cite{ETMC2,ETMC3} and $N_f=3$ \cite{HPQCD2,FNAL+MILC2}.}
\end{figure}

We now perform a bootstrap analysis of $f_{B_s}$ combining all OPE
uncertainties. We assume Gaussian distributions~of the OPE
parameters (quark masses, condensates) with corresponding errors.
The renormalization scale $\mu$ is assumed~to be uniformly
distributed in the interval $2\le \mu\;(\mbox{GeV})\le 8$. Making
use of the range $m_b=(4.163\pm 0.016)\;\mbox{GeV}$ \cite{mb}~one
has $f_{B_s} = (262.0 \pm 18.1_{\rm (OPE)} \pm 2.9_{\rm (syst)})\;
{\rm MeV}$, while the range (\ref{mb}), $m_b=(4.245\pm
0.025)\;\mbox{GeV}$, leads to
\begin{eqnarray}
\label{fBs} f_{B_s} = (232.5 \pm 18.6_{\rm (OPE)} \pm 2.4_{\rm
(syst)})\; {\rm MeV}.
\end{eqnarray}
Figure~\ref{Plot:fBs_bootstrap}b compares our results with recent
lattice determinations.

\subsection{\boldmath $f_{B_s}/f_{B}$}
For the ratio of the decay constants of beauty mesons, we find
\begin{eqnarray}
\label{ratioB}
f_{B_s}/f_B = 1.203\pm 0.020_{(\rm OPE)}\pm 0.007_{(\rm syst)},
\end{eqnarray}
which has to be compared with the lattice results $f_{B_s}/f_B =
1.27\pm 0.05 $ at $N_f=2$ \cite{ETMC3} and $f_{B_s}/f_B = 1.226\pm
0.026$~\cite{HPQCD2} and $f_{B_s}/f_B = 1.245\pm 0.043$
\cite{FNAL+MILC2} at $N_f=3$. Similar to the charmed ratio
(\ref{ratioD}), the error in (\ref{ratioB}) comes mainly from~the
uncertainty in the ratio of quark condensates: $\langle \bar s
s\rangle/\langle \bar q q\rangle=0.8\pm 0.3$.

\subsection{\boldmath Double ratio $(f_{B_s}/f_{B})/(f_{D_s}/f_{D})$}
The double ratio of the decay constants --- the Grinstein ratio
$R_1$ \cite{grinstein} --- is practically free from OPE
uncertainties. We obtain the particularly accurate value
\begin{eqnarray}
\label{doubleratio}
\frac{f_{B_s}/f_B}{f_{D_s}/f_D} -1 = 0.013\pm 0.011_{(\rm syst)},
\end{eqnarray}
which is consistent with the lattice determination
$(f_{B_s}/f_B)/(f_{D_s}/f_D) = 0.018\pm 0.006\pm 0.010$
\cite{double_lat}.

\section{Summary and conclusions}\label{Sec:S&C}
In summary, we performed a detailed analysis of the extraction of
the decay constants of pseudoscalar heavy mesons from the
correlator of pseudoscalar currents. Particular emphasis was laid
on the investigation of the uncertainties~in the extracted values
of the decay constants: namely, on the OPE uncertainty related to
the not precisely known~QCD parameters, and on the intrinsic
uncertainty of the method related to the limited accuracy of the
extraction procedure. According to our recent findings, the
accuracy of the sum-rule estimates may be considerably improved
and the intrinsic uncertainties in hadron parameters may be probed
by studying systematically the Borel-parameter dependence of the
effective continuum thresholds; the parameters of these effective
thresholds may be fixed by minimizing the deviation~of the dual
mass from the known meson mass in the Borel window. In the present
work, this strategy has been applied~to the decay constants of
heavy mesons. Our main results are as follows:

(i) We obtain the following estimates for the decay constants of
the charmed $D$ and $D_s$ mesons:
\begin{eqnarray}
\label{fD_final} f_{D}&=& (206.2 \pm 7.3_{\rm (OPE)} \pm 5.1_{\rm
(syst)})\; \mbox{MeV}, \\ \label{fDs_final} f_{D_s}&=& (245.3 \pm
15.7_{\rm (OPE)} \pm 4.5_{\rm (syst)})\; \mbox{MeV}.
\end{eqnarray}
We would like to point out that we provide both the OPE
uncertainties and the intrinsic (systematic) uncertainty~of the
method of sum rules related to the limited accuracy of the
extraction procedure. In the case of $f_D$, the latter~turns out
to be of the same order of magnitude as the OPE uncertainty.
Noteworthy, assuming $\tau$-independence of the effective
continuum threshold leads to the substantially lower
decay-constant range $f_D{(n=0)} = (181.3 \pm 7.4_{\rm (OPE)})\;
\mbox{MeV}$, which differs by almost three times the OPE
uncertainty from our result (\ref{fD_final}) found from a
$\tau$-dependent effective threshold. The ratio of the
charmed-meson decay constants (\ref{fDs_final}) and
(\ref{fD_final}) is
\begin{eqnarray}
\label{ratioD_final}
f_{D_s}/f_D = 1.193\pm 0.025_{(\rm OPE)}\pm 0.007_{(\rm syst)}.
\end{eqnarray}

(ii) The decay constants of the $B$ and $B_s$ mesons are very
sensitive to the precise value of $m_b$. Using the PDG~range of
$m_b$ does not allow us to obtain a reasonable estimate. For the
very narrow range $\overline{m}_b(\overline{m}_b)=(4.163\pm
0.0016)\,\mbox{GeV}$~\cite{mb}, our analysis gives $f_{B} = (225.6
\pm 11.3_{\rm (OPE)} \pm 2.2_{\rm (syst)})\; {\rm MeV}$ and $
f_{B_s} = (262.0 \pm 18.1_{\rm (OPE)} \pm 2.9_{\rm (syst)})\; {\rm
MeV}$. We observe some tension between the above sum-rule result
for $f_B$ and the average of recent lattice calculations
\cite{ETMC2,ETMC3,HPQCD2,FNAL+MILC2},~namely, $f_B^{(\rm
lattice)}=193 \pm 13$ MeV.

We emphasize that the observed strong sensitivity of $f_B$ to the
precise value of $m_b$ provides an interesting alternative way of
obtaining $m_b$ from the analysis of the decay constant: Using the
lattice average for $f_B$ as input yields the~rather accurate
estimate for the $b$-quark mass
\begin{eqnarray}
\label{mb_final} \overline{m}_b(\overline{m}_b)=(4.245\pm 0.025)\;
\mbox{GeV}.
\end{eqnarray}
This new range of $m_b$ corresponds to
\begin{eqnarray}
\label{fB_final} f_{B} = (193.4 \pm 12.3_{\rm (OPE)} \pm 4.3_{\rm
(syst)})\; {\rm MeV}
\end{eqnarray}
and yields
\begin{eqnarray}
\label{fBs_final} f_{B_s} = (232.5 \pm 18.6_{\rm (OPE)} \pm
2.4_{\rm (syst)})\; {\rm MeV}.
\end{eqnarray}
For the ratio of the decay constants (\ref{fBs_final}) and
(\ref{fB_final}) we get
\begin{eqnarray}
\label{ratioB_final}
f_{B_s}/f_B = 1.203\pm 0.020_{(\rm OPE)}\pm 0.007_{(\rm syst)},
\end{eqnarray}

(iii) The double (Grinstein) ratio of the decay constants,
\begin{eqnarray}
\frac{f_{B_s}/f_B}{f_{D_s}/f_D} -1 = 0.013\pm 0.011_{(\rm syst)},
\end{eqnarray}
is practically free from OPE uncertainties and, consequently, may
be predicted with rather high accuracy.

\acknowledgments We are indebted to Matthias Jamin for providing
us with his Mathematica code for the calculation of the two-point
function. DM expresses his gratitude to the Institute of
Theoretical Physics of the Heidelberg University for hospitality
during his visit to Heidelberg, where this work was started. DM
was supported, in part, by: the Austrian~Science Fund (FWF) under
project P20573, the Alexander von Humboldt-Stiftung, the Federal
Agency for Science and Innovation of Russian Federation under
state contract 02.740.11.0244, and EU Contract
No.~MRTN-CT-2006-035482~``FLAVIAnet''.

\end{document}